\documentclass[aps,twocolumn]{revtex4}

\usepackage{graphicx}
\usepackage{natbib}

\usepackage{graphics}
\usepackage{epsfig}
\usepackage{bm}
\usepackage{amsmath,amssymb}
\usepackage{stmaryrd}

\def\wt#1{\widetilde{#1}}
\def\vb#1{\mbox{\boldmath$#1$}}
\def\pd#1#2{\frac{\partial #1}{\partial #2}}

\def\wh#1{\widehat{#1}}
\def\bdot{\,\vb{\cdot}\,}
\def\btimes{\,\vb{\times}\,}

\def\bhat{\wh{{\sf b}}}
\def\cal#1{\mathcal{#1}}

\def\exd{{\sf d}}
\def\bhat{\wh{{\sf b}}}

\newcommand{\bc}{\begin{center}}
\newcommand{\ec}{\end{center}}
\newcommand{\bt}{\begin{tabbing}}
\newcommand{\et}{\end{tabbing}}
\newcommand{\be}{\begin{equation}}
\newcommand{\ee}{\end{equation}}
\newcommand{\ba}{\begin{eqnarray}}
\newcommand{\ea}{\end{eqnarray}}

\begin{document}

\title{Variational Formulation of Higher-order Guiding-center Vlasov-Maxwell Theory}

\author{Alain J.~Brizard$^{a}$}
\affiliation{Department of Physics, Saint Michael's College, Colchester, VT 05439, USA \\ $^{a}$Author to whom correspondence should be addressed: abrizard@smcvt.edu}

\begin{abstract}
Extended guiding-center Vlasov-Maxwell equations are derived under the assumption of time-dependent and inhomogeneous electric and magnetic fields that obey the standard guiding-center space-time-scale orderings. The guiding-center Vlasov-Maxwell equations are derived up to second order, which contain dipole and quadrupole contributions to the guiding-center polarization and magnetization that include finite-Larmor-radius corrections. Exact energy-momentum conservation laws are derived from the variational formulation of these higher-order guiding-center Vlasov-Maxwell equations. 
\end{abstract}

\date{\today}

\maketitle

\section{Introduction}

The adiabatic invariance of the magnetic moment along a charged-particle orbit (with mass $m$ and charge $e$) in a nonuniform magnetic field ${\bf B} = B\,\bhat$ plays a crucial role in our understanding of the physical basis of the spatial confinement of a magnetized plasma over long time scales \citep{Hazeltine_Meiss_2003,Helander_2014}. 

In guiding-center theory \cite{Northrop1963,Cary_Brizard_2009}, the mathematical construction of the magnetic moment relies on the space-time scales $(L_{B},\omega^{-1})$ of the confining magnetic field to be long compared to the characteristic gyroradius $\rho$ and the gyroperiod $\Omega^{-1} = mc/eB$, respectively, leading to the small dimensionless ordering parameter \cite{Cary_Brizard_2009}
\begin{equation}
\epsilon_{B} \;\equiv\; \rho/L_{B} \;\sim\; \omega/\Omega \;\ll\; 1. 
\label{eq:ordering}
\end{equation}
The traditional derivation of guiding-center theory relies on the existence of an ordering parameter defined by the dimensional mass-to-charge ratio $m/e$ \cite{Northrop1963,Kruskal_1965}, so that the gyroperiod $\Omega^{-1} \propto m/e$ is assumed to be the shortest time scale. For practical applications in perturbation theory \cite{RGL_1981}, however, it is convenient to replace the ratio $m/e$ with $\epsilon\,m/e = (\epsilon\, m)/e = m/(\epsilon^{-1}e)$, so that the dimensionless ordering parameter 
$\epsilon \ll 1$ can either be viewed as a {\it renormalization} of the particle's mass $m \rightarrow \epsilon\,m$ or the particle's charge $e \rightarrow e/\epsilon$ and it is quite common to assume $\epsilon \sim \epsilon_{B}$. It is noteworthy, however, that the guiding-center approximation is still valid when the spatial ordering $\rho/L_{B} \lesssim 1$ is not extremely small (as recently demonstrated in Ref.~\cite{Brizard_2017}).

In earlier derivations of the guiding-center Vlasov equation (sometimes referred to as the drift-kinetic equation), a recursive solution of the Vlasov kinetic equation \cite{CGL_1956,Frieman_1966,Hazeltine_1973,Kulsrud_1983} led to the drift-kinetic equation through an expansion of the Vlasov distribution function $f = f_{0} + \epsilon\,f_{1} + \epsilon^{2}\,f_{2} + \cdots$, which yielded functional solutions $f_{n}[f_{0}]$ ($n \geq 1$) in terms of the lowest-order gyroangle-independent solution $f_{0}$. It was later shown that this recursive derivation is completely analogous to the Lie-transform derivation of the guiding-center Vlasov equation \cite{Brizard_Mishchenko_2009} in the electrostatic limit. It is precisely because of their power and simplicity that Lie-transform perturbation methods are used here to derive higher-order guiding-center equations of motion. We also note that, when Lie-transform perturbation methods are combined with variational formulations for the reduced Vlasov-Maxwell equations \cite{Brizard_2008}, the reduced Maxwell equations naturally incorporate reduced polarization and magnetization effects while exact energy-momentum conservation laws are derived by Noether method \cite{Brizard_2000_prl}.

\subsection{Motivation}

The inclusion of nonuniform, time-dependent electric fields in guiding-center theory has a long and rich history in plasma physics  \cite{Northrop1963,Cary_Brizard_2009}. In the present work, we use the standard ordering \cite{Kulsrud_1983} for the electric field 
${\bf E} = {\bf E}_{\bot} + \epsilon\,E_{\|}\,\bhat$, where the parallel component $E_{\|} \rightarrow \epsilon\,E_{\|}$ of the electric field is considered small compared to the perpendicular components ${\bf E}_{\bot}$. In addition, we assume that the $E\times B$ velocity $c|{\bf E}_{\bot}|/B$ is comparable to particle's thermal velocity.

The ability of electric fields to fundamentally modify the magnetic geometry that confines a laboratory plasma (e.g., by creating transport barriers \cite{Hahm_1996,Itoh_1996,Terry_2000,Ida_2018,Dif-Pradalier_2022} or in setting up rotating mirror magnetic geometries \cite{White_2018}) motivates the need to construct a guiding-center Vlasov-Maxwell theory that includes self-consistent time-dependent electric and magnetic fields, in which the transfer of energy and momentum between the electromagnetic fields and the confined plasma play a crucial role. 

In addition, for many situations of practical interest, the presence of a strong electric field is associated with strong plasma flows with steep sheared rotation profiles for which second-order effects (including finite-Larmor-radius effects), which must be included in a self-consistent guiding-center theory \cite{Hahm_1996,Chang_2004,Lanthaler_2019,Frei_2020,Joseph_2021,Dif-Pradalier_2022}. Guiding-center equations of motion with second-order corrections in the presence of time-independent electric and magnetic fields were derived using Lie-transform perturbation method by Brizard \cite{Brizard_1995} and Hahm \cite{Hahm_1996}, following the earlier work of Littlejohn \cite{RGL_1981}. More recently, these perturbation methods were also used by Miyato {\it et al.} \cite{Miyato_Scott_2009} and Madsen \cite{Madsen_2010}, who derived self-consistent guiding-center Vlasov-Maxwell equations that included guiding-center polarization and magnetization effects. Not all second-order effects were included in these models, however, and it is the purpose of the present work to derive a more complete higher-order guiding-center Vlasov-Maxwell theory, based on the Lie-transform perturbation derivation of higher-order guiding-center Lagrangian dynamics \cite{Brizard_2023_arxiv} for the case of time-dependent, nonuniform electric and magnetic fields that satisfy the guiding-center ordering \eqref{eq:ordering}.

We note that a key difference between guiding-center Vlasov-Maxwell models considered here and gyrokinetic Vlasov-Maxwell models considered elsewhere (see Ref.~\cite{Brizard_Hahm_2007} for a review), is that the electromagnetic fields $({\bf E},{\bf B})$ considered here are not separated into time-independent weakly-nonuniform background fields and time-dependent fluctuating (i.e., turbulent) fields that may possibly have short spatial scales (satisfying the gyrokinetic ordering \cite{Brizard_Hahm_2007}). 
Hence, the guiding-center Vlasov-Maxwell energy and momentum are exactly conserved despite the fact that guiding-center Vlasov-Maxwell fields are time-dependent and nonuniform, with space-time scales \eqref{eq:ordering} that satisfy the guiding-center orderings \cite{Cary_Brizard_2009}.

\subsection{The need for higher-order guiding-center theory}

The previous variational derivations of a self-consistent guiding-center Vlasov-Maxwell model have been carried out up to first order in the dimensionless ordering parameter $\epsilon$ 
\cite{Pfirsch_1984,Pfirsch_Morrison_1985,Ye_Kaufman_1992,Brizard_Tronci_2016,Sugama_2016,Brizard_2021}. In Ref.~\cite{Brizard_2021}, for example, the Hamiltonian structure of the first-order guiding-center Vlasov-Maxwell equations was given in terms of a guiding-center Hamiltonian functional and a functional bracket that satisfies the Jacobi property. The variational and Hamiltonian structures of the guiding-center Vlasov-Maxwell equations may prove useful in the implementation of structure-preserving numerical algorithms \cite{Ellison_2015,Morrison_2017,Kraus_2017,Xiao_Qin_Liu_2018,Ellison_2018} .

Recently, second-order terms in guiding-center Hamiltonian theory (in the absence of an electric field) were shown to be crucial \cite{Brizard_Hodgeman_2023} in assessing the validity of the guiding-center representation in determining whether guiding-center orbits were numerically faithful to the particle orbits in axisymmetric magnetic geometries, which partially confirmed earlier numerical studies in axisymmetric tokamak plasmas \cite{Belova_2003}. In particular, it was shown that a second-order correction associated with guiding-center polarization \cite{Kaufman_1986,Brizard_2013,Tronko_Brizard_2015} was needed in order to obtain faithful guiding-center orbits. 

Indeed, without the inclusion of second-order effects, it was shown that, within a few bounce periods after leaving the same physical point in particle phase space, a first-order guiding-center orbit deviated noticeably from its associated particle orbit, while a second-order guiding-center orbit followed the particle orbit to a high degree of precision \cite{Brizard_Hodgeman_2023}. In addition, as initially reported by Belova {\it et al.} \cite{Belova_2003}, the guiding-center Hamiltonian formulation \cite{Tronko_Brizard_2015,Brizard_Hodgeman_2023} is a faithful representation of the particle toroidal angular momentum, which is an exact particle constant of motion in an axisymmetric magnetic field, only if second-order effects are included. See additional comments included in Sec.~\ref{sec:P_phi} regarding the faithfulness of the guiding-center representation.

From a purely theoretical point of view, it is therefore interesting to derive higher-order guiding-center Vlasov-Maxwell equations with accurate expressions for the guiding-center polarization and magnetization, which include finite-Larmor-radius (FLR) corrections. Through the application of the Noether method  \cite{Brizard_2000_prl}, we will also be able to explore how these higher-order effects modify the guiding-center energy-momentum conservation laws, e.g., how the Chew-Goldberger-Low (CGL) pressure tensor \cite{CGL_1956,Frieman_1966,Hazeltine_1973,Kulsrud_1983} is modifed [see Eq.~\eqref{eq:Tgc_sym}].

\subsection{Organization}

The remainder of the present work is organized as follows. In Sec.~\ref{sec:gc_Ham}, the extended guiding-center Hamilton equations of motion are derived in terms of the extended guiding-center Hamiltonian and the extended guiding-center Poisson bracket (derived by Lie-transform perturbation method in Ref.~\cite{Brizard_2023_arxiv}), while the higher-order guiding-center Vlasov-Maxwell equations are derived from an Eulerian variational principle \cite{Brizard_2000_prl} in Sec.~\ref{sec:gc_var}, from which guiding-center polarization and magnetization are derived with FLR corrections. Here, our guiding-center variational principle expressly imposes plasma quasineutrality by omitting the electric energy density from the guiding-center Lagrangian density \cite{Newcomb_1962,Grad_1966,Scott_2010,Tronko_2018,McMillan_2023}, which also removes the displacement current $(c^{-1}\partial{\bf E}/\partial t)$ from Maxwell's equation. 

In Sec.~\ref{sec:gc_Noether}, the exact energy-momentum conservation laws are derived from the guiding-center Noether equation obtained from the guiding-center Eulerian variational principle. The symmetry properties of the guiding-center stress tensor are also briefly discussed, and the guiding-center angular-momentum conservation law is derived at the lowest order, while a more extensive discussion of the asymmetry of the guiding-center stress tensor  at higher order is left for future work.

Finally, in App.~\ref{sec:gyrogauge}, we present a brief discussion of the concept of gyrogauge invariance \cite{RGL_1981,RGL_1983,RGL_1988}, which states that our guiding-center equations of motion should not only be independent of the gyroangle but also how the angle is measured. Gyrogauge terms appear at second order in the guiding-center perturbation analysis and were shown to be crucial in previous works \cite{Brizard_Hodgeman_2023,Belova_2003}. In App.~\ref{sec:linear_gcVM}, we show how the standard linear finite-beta electromagnetic gyrokinetic equations \cite{Brizard_Hahm_2007} form a subset of our guiding-center Vlasov-Maxwell equations in the limit of a uniform background magnetized plasma. In particular, we show how well-known spurious high-frequency modes found in electrostatic gyrokinetic models \cite{Lee_1983,Dubin_1983,Lee_1987,Krommes_1993} disappear when electromagnetic effects are included.

\section{\label{sec:gc_Ham}Guiding-center Hamiltonian Dynamics} 

In this Section, we make use of the results of the Lie-transform perturbation analysis presented in Ref.~\cite{Brizard_2023_arxiv}, that yield the guiding-center phase-space extended one-form (expanded in terms of the mass-renormalization ordering $m \rightarrow \epsilon\,m$)
\begin{eqnarray}
\Gamma_{\rm gc} &=& \left( \frac{e}{c}\,{\bf A} \;+\; \epsilon\,\vb{\Pi}_{\rm gc}\right)\bdot\exd{\bf X} \;-\;  W\;\exd t \nonumber \\
 &&+\; \epsilon^{2}\,J \left(\exd\theta - {\bf R}\bdot\exd{\bf X} - {\cal S}\,\exd t\right) \nonumber \\
 &\equiv& \frac{e}{c}\;{\bf A}^{*}\bdot\exd{\bf X} \;+\; \epsilon^{2}\,J\;\exd\theta \;-\; W^{*}\,\exd t,
\label{eq:Gamma_gc_primitive}
\end{eqnarray}
where we introduced the definitions
\begin{eqnarray}
 \frac{e}{c}\,{\bf A}^{*} &\equiv& \frac{e}{c}\,{\bf A} + \epsilon\,{\bf P}_{0} - \epsilon^{2}\,J\,{\bf R}^{*}, \label{eq:Astar_def} \\
W^{*} &\equiv& W \;+\; \epsilon^{2}\,J\,{\cal S}. \label{eq:Wstar_def}
\end{eqnarray}
where 
\begin{equation}
{\bf P}_{0} \;=\; P_{\|}\,\bhat \;+\; {\bf E}\btimes e\bhat/\Omega
\label{eq:Pi_gc_def}
\end{equation} 
is the lowest guiding-center momentum and ${\bf R}^{*} \equiv {\bf R} + \frac{1}{2}\,\nabla\btimes\bhat$ includes the second-order polarization correction $\vb{\Pi}_{\rm pol} \equiv -\,\frac{1}{2}\,J\,\nabla\btimes\bhat$ introduced by Tronko and Brizard \cite{Tronko_Brizard_2015} in order to obtain an exact Lie-transform derivation of the guiding-center polarization derived by Kaufman \cite{Kaufman_1986}. In addition, the presence of the gyrogauge fields $({\cal S},{\bf R})$ in Eqs.~\eqref{eq:Gamma_gc_primitive}-\eqref{eq:Wstar_def} guarantee gyrogauge invariance \cite{RGL_1981,RGL_1988}. 

The extended guiding-center Hamiltonian, on the other hand, is expressed as
\begin{equation}
{\cal H}_{\rm gc} \;=\; e\,\Phi \;+\; \epsilon\,K_{\rm gc} \;+\; \epsilon^{2}\;J\,{\cal S}^{*} \;-\; W^{*} \;\equiv\; e\,\Phi^{*} \;-\; W^{*},
\end{equation}
 where the guiding-center kinetic energy in the drifting frame is
 \begin{eqnarray}
K_{\rm gc} &=& \mu\,B \;+\; \frac{P_{\|}^{2}}{2m} \;+\; \frac{m}{2}\,|{\bf u}_{\rm E}|^{2}  \label{eq:K_gc_def} \\
{\cal S}^{*} &=& {\cal S} \;-\; \nabla\bdot\left( \frac{\bhat}{2}\btimes{\bf u}_{\rm E}\right),
\end{eqnarray}
which includes the second-order FLR correction to the electrostatic potential energy $e\,\Phi$.  This FLR correction may be decomposed as 
  \[ -\;\nabla\bdot\left( \frac{1}{2}J\bhat\btimes{\bf u}_{\rm E}\right) \;=\; -\;\frac{J}{2}\left( \nabla\btimes\bhat\bdot{\bf u}_{\rm E} \;-\frac{}{} \bhat\bdot\nabla\btimes{\bf u}_{\rm E}\right), \]
  which includes the standard second-order guiding-center Hamiltonian $\frac{1}{2}\,J\,\bhat\bdot\nabla\btimes{\bf u}_{\rm E}$ \cite{RGL_1981,RGL_1983,Hahm_1996,Madsen_2010}, and the new guiding-center polarization correction 
  $\vb{\Pi}_{\rm pol}\bdot{\bf u}_{\rm E}$, which is ignored by these previous works.
  
  In the remainder of the paper, we will remove the ordering parameter $\epsilon$ and return to the physical mass $\epsilon\,m \rightarrow m$, while we may occasionally refer to this mass-renormalization ordering in what follows.

\subsection{Extended guiding-center Poisson bracket}

The extended guiding-center Poisson bracket $\{\;,\; \}_{\rm gc}$ is obtained by, first, constructing an 8$\times$8 matrix out of the components of the extended guiding-center Lagrange two-form $\vb{\omega}_{\rm gc} = \exd\Gamma_{\rm gc}$ and, then, invert this matrix to obtain the extended guiding-center Poisson matrix, whose components are the fundamental brackets $\{Z^{\alpha}, Z^{\beta}\}_{\rm gc}$. 

From these components, we obtain the  extended guiding-center Poisson bracket \cite{Brizard_2023_arxiv}
\begin{eqnarray}
\{{\cal F}, {\cal G}\}_{\rm gc} &=& \frac{{\bf B}^{*}}{B_{\|}^{*}}\bdot\left(\nabla^{*}{\cal F}\,\pd{\cal G}{P_{\|}} - \pd{\cal F}{P_{\|}}\,\nabla^{*}{\cal G}\right) \nonumber \\
&&-\; \frac{c\bhat}{eB_{\|}^{*}}\bdot\nabla^{*}{\cal F}\btimes\nabla^{*}{\cal G} \nonumber \\
 &&+\; \left(\pd{\cal F}{\theta}\,\pd{\cal G}{J} - \pd{\cal F}{J}\,\pd{\cal G}{\theta}\right) \nonumber \\
   &&+\; \left(\pd{\cal F}{W}\,\frac{\partial^{*}{\cal G}}{\partial t} - \frac{\partial^{*}{\cal G}}{\partial t}\,\pd{\cal G}{W}\right), 
   \label{eq:gcPB_ext}
 \end{eqnarray}
 where 
 \begin{equation}
 \frac{e}{c}\,{\bf B}^{*} \;=\; \frac{e}{c}\,{\bf B} \;+\; \nabla\btimes{\bf P}_{0} \;-\; J\,\nabla\btimes{\bf R}^{*}, \label{eq:B_star}
 \end{equation}
and the guiding-center Jacobian is ${\cal J}_{\rm gc} = (e/c)\,B_{\|}^{*}$, where $B_{\|}^{*} \equiv \bhat\bdot{\bf B}^{*}$. In addition, we introduced the definitions
 \begin{eqnarray}
 \frac{\partial^{*}}{\partial t} &\equiv& \pd{}{t} \;+\; {\cal S}\;\pd{}{\theta}, \label{eq:t_star} \\
 \nabla^{*} &\equiv& \nabla \;+\; {\bf R}^{*}\;\pd{}{\theta} \;-\; \left( \frac{e}{c}\pd{{\bf A}^{*}}{t} + J\,\nabla{\cal S}\right)\pd{}{W}, \label{eq:grad_star}
 \end{eqnarray}
We note that the Poisson bracket \eqref{eq:gcPB_ext} can be expressed in divergence form as
\begin{equation}
\{{\cal F}, {\cal G}\}_{\rm gc} \;=\; \frac{1}{B_{\|}^{*}}\pd{}{Z^{\alpha}}\left(B_{\|}^{*}\frac{}{}{\cal F}\;\left\{ Z^{\alpha},\; {\cal G}\right\}_{\rm gc}\right),
\label{eq:gcPB_div}
\end{equation}
and that it automatically satisfies the Jacobi identity
 \begin{equation}
 \left\{{\cal F},\frac{}{}\{{\cal G},{\cal K}\}\right\} +  \left\{{\cal G},\frac{}{}\{{\cal K},{\cal F}\}\right\} +  \left\{{\cal K},\frac{}{}\{{\cal F},{\cal G}\}\right\} = 0.
 \end{equation}
Next, we note that the operators \eqref{eq:t_star} and \eqref{eq:grad_star} contain the gyrogauge-invariant combinations $\partial/\partial t + {\cal S}\,\partial/\partial\theta$ and $\nabla + {\bf R}\;\partial/\partial\theta$, while Eqs.~\eqref{eq:B_star} and \eqref{eq:grad_star} include the gyrogauge-independent vector fields (see App.~\ref{sec:gyrogauge})
 \begin{eqnarray}
 \nabla\btimes{\bf R} &=& -\,\frac{1}{2}\,\epsilon_{ijk}\,b^{i}\;\nabla b^{j}\btimes\nabla b^{k}, \label{eq:R_vector}\\
 \nabla{\cal S} \;-\; \pd{\bf R}{t} &=& -\,\nabla\bhat\btimes\bhat\bdot\pd{\bhat}{t}, \label{eq:S_scalar}
 \end{eqnarray}
 where $\epsilon_{ijk}$ denotes the completely-antisymmetric Levi-Civita tensor.

\subsection{Guiding-center Hamilton equations}

 The guiding-center Hamilton equations include the guiding-center velocity
 \begin{equation}
 \dot{\bf X} \;\equiv\; \{{\bf X},\;{\cal H}_{\rm gc}\}_{\rm gc} \;=\; \frac{P_{\|}}{m}\;\frac{{\bf B}^{*}}{B_{\|}^{*}} \;+\; {\bf E}^{*}\btimes\frac{c\bhat}{B_{\|}^{*}}, 
 \label{eq:Xgc_dot}
 \end{equation}
 where 
 \begin{equation}
 \bhat\bdot\dot{\bf X} \;=\; \partial{\cal H}_{\rm gc}/\partial P_{\|} \;=\; P_{\|}/m 
 \label{eq:ELgc_P}
 \end{equation}
 defines the parallel guiding-center velocity, the guiding-center parallel force
 \begin{equation}
 \dot{P}_{\|} \;\equiv\; \{P_{\|},\;{\cal H}_{\rm gc}\}_{\rm gc} \;=\; e\,{\bf E}^{*}\bdot\frac{{\bf B}^{*}}{B_{\|}^{*}},
 \label{eq:Pgc_dot}
 \end{equation}
 where the modified electric field is represented as
 \begin{eqnarray}
 e\,{\bf E}^{*} &=& -\;e\,\nabla\Phi^{*} \;-\; \frac{e}{c}\,\pd{{\bf A}^{*}}{t} \nonumber \\
  &=& e\,{\bf E} - \pd{{\bf P}_{0}}{t} - \nabla K_{\rm gc} + J \left(\pd{{\bf R}^{*}}{t} - \nabla{\cal S}^{*}\right), 
  \label{eq:E_star}
  \end{eqnarray}
 and the gyroangle angular velocity
 \begin{eqnarray}
 \dot{\theta} &\equiv& \{\theta,\; {\cal H}_{\rm gc}\}_{\rm gc} \;=\; \pd{K_{\rm gc}}{J} \;+\; {\cal S}^{*} \;+\; \dot{\bf X}\bdot{\bf R}^{*} \nonumber \\
 &=& \Omega \;-\; \frac{1}{2}\;\nabla\bdot\left(\bhat\btimes{\bf u}_{\rm E}\right) \;+\; {\cal S} \;+\; \dot{\bf X}\bdot{\bf R}^{*}. 
 \label{eq:theta_dot_gc}
  \end{eqnarray}
  We note that the guiding-center Hamilton equations \eqref{eq:Xgc_dot} and \eqref{eq:Pgc_dot} can also be derived from the guiding-center Lagrangian \cite{Brizard_2023_arxiv}
  \begin{equation}
  L_{\rm gc} \;=\; \frac{e}{c}\,{\bf A}^{*}\bdot\dot{\bf X} \;+\; J\,\dot{\theta} \;-\; e\,\Phi^{*}.
  \label{eq:Lag_gc_def}
   \end{equation}
   as the guiding-center Euler-Lagrange equation
  \begin{equation}
  \dot{P}_{\|}\,\bhat \;=\; e\,{\bf E}^{*} \;+\; \frac{e}{c}\,\dot{\bf X}\btimes{\bf B}^{*}
  \label{eq:ELgc_X}
  \end{equation}
  together with Eq.~\eqref{eq:ELgc_P}. If we omit the terms $({\bf R}^{*},{\cal S}^{*})$ in Eqs.~\eqref{eq:B_star} and \eqref{eq:E_star}, we also note that the guiding-center velocity can be expressed as
  \begin{equation}
  \dot{\bf X} \;=\; \frac{{\bf P}_{0}}{m} \;+\;  \frac{c\bhat}{eB_{\|}^{*}}\btimes\left(\frac{d_{0}{\bf P}_{0}}{dt} \;+\; \mu\,\nabla B\right),
  \label{eq:xdot_P0}
  \end{equation}
  where ${\bf P}_{0} \equiv P_{\|}\,\bhat + {\bf E}\btimes e\bhat/\Omega$ denotes the lowest-order guiding-center momentum and $d_{0}/dt \equiv \partial/\partial t + ({\bf P}_{0}/m)\bdot\nabla$, while the higher-order corrections to the guiding-center velocity include the standard magnetic-drift velocity $(c\bhat/eB_{\|}^{*})\btimes[(P_{\|}^{2}/m)\bhat\bdot\nabla\bhat + \mu\,\nabla B]$, as well as the total polarization drift $(mc\bhat/eB_{\|}^{*})\btimes d_{0}{\bf u}_{\rm E}/dt$.

   Finally, we note that the guiding-center Jacobian ${\cal J}_{\rm gc} = (e/c)\,B_{\|}^{*}$ satisfies the guiding-center Liouville equation
 \begin{equation}
 \pd{B_{\|}^{*}}{t} \;=\; -\;\nabla\bdot\left(B_{\|}^{*}\;\dot{\bf X}\right) \;-\; \pd{}{P_{\|}}\left(B_{\|}^{*}\;\dot{P}_{\|}\right),
 \label{eq:gc_Liouville}
 \end{equation} 
 where
 \begin{eqnarray*} 
 \nabla\bdot\left(B_{\|}^{*}\;\dot{\bf X}\right) &=& \nabla\btimes{\bf E}^{*}\bdot c\bhat - e\,{\bf E}^{*}\bdot\frac{c}{e}\nabla\btimes\bhat \nonumber \\
  &=& -\,\bhat\bdot\pd{{\bf B}^{*}}{t} - e\,{\bf E}^{*}\bdot\pd{{\bf B}^{*}}{P_{\|}}
 \end{eqnarray*}
 and
 \begin{eqnarray*}
 \pd{}{P_{\|}}\left(B_{\|}^{*}\;\dot{P}_{\|}\right) &=& e\,{\bf E}^{*}\bdot\pd{{\bf B}^{*}}{P_{\|}} + {\bf B}^{*}\bdot e\pd{{\bf E}^{*}}{P_{\|}} \\
  &=& e\,{\bf E}^{*}\bdot\pd{{\bf B}^{*}}{P_{\|}} - {\bf B}^{*}\bdot \pd{\bhat}{t},
 \end{eqnarray*}
 where we made use of the modified Faraday's law $\partial{\bf B}^{*}/\partial t = -\,c\,\nabla\btimes{\bf E}^{*}$.

  \subsection{\label{sec:P_phi} Guiding-center canonical toroidal angular momentum}
  
   By applying Noether's method \cite{Brizard_2015} on the guiding-center Lagrangian \eqref{eq:Lag_gc_def}, we immediately find that, in the case of an axisymmetric magnetized plasma (i.e., $\partial L_{\rm gc}/\partial\phi \equiv 0$), the guiding-center canonical toroidal angular momentum
   \begin{eqnarray}
   P_{\phi} &\equiv& \pd{L_{\rm gc}}{\dot{\phi}} \;=\; \frac{e}{c}\,{\bf A}^{*}\bdot\pd{\bf X}{\phi} \nonumber \\
    &=& \frac{e}{c}\,\left(A_{\phi} \;-\; \vb{\rho}_{\rm E}\bdot\nabla A_{\phi}\right) \;+\; P_{\|}\,b_{\phi}  \nonumber \\
     &&-\; J \left( b_{z} \;+\; \frac{1}{2}\,\nabla\btimes\bhat\bdot\pd{\bf X}{\phi} \right),
    \label{eq:P_phi}
   \end{eqnarray}
   is a constant of the guiding-center motion, where $(A_{\phi},b_{\phi})$ denote covariant toroidal components and we used the identity ${\bf R}\bdot\partial{\bf X}/\partial\phi \equiv b_{z}$ \cite{RGL_1983}, which makes $P_{\phi}$ gyrogauge invariant. Here, we used the identity ${\bf B}\btimes\partial{\bf X}/\partial\phi = -\,\nabla A_{\phi}$ and we introduced the electric polarization displacement $\vb{\rho}_{\rm E} \equiv (c/B\Omega)\,{\bf E}_{\bot}$ in writing $\vb{\Pi}_{\rm E}\bdot\partial{\bf X}/\partial\phi = -\,(e/c)\,
   \vb{\rho}_{\rm E}\bdot\nabla A_{\phi}$. 
   
   We note that the issue of the faithfulness of the guiding-center representation, which was initiated by Belova {\it et al.}~\cite{Belova_2003} in their numerical studies of particle and guiding-center orbits of energetic ions in axisymmetric tokamak geometry (in the absence of an electric field), focussed on whether the guiding-center pull-back of the guiding-center canonical toroidal angular momentum ${\sf T}_{\rm gc}P_{\phi}$ accurately describes the particle canonical toroidal angular momentum $p_{\phi} \equiv 
   \partial L/\partial\dot{\phi}$ defined in terms of the particle Lagrangian $L$.
   
   Belova {\it et al.}~\cite{Belova_2003} noted that the numerical plot of ${\sf T}_{\rm gc}P_{\phi}$ for energetic ions (with $\epsilon_{B} \lesssim 0.2$) shows excellent invariance properties equal to the true particle invariant $p_{\phi}$ (within numerical accuracy) only when second-order $(\epsilon^{2})$ corrections are included in the guiding-center canonical toroidal angular momentum \eqref{eq:P_phi} and the guiding-center pull-back operator ${\sf T}_{\rm gc}$ (defined in terms of generators of the guiding-center phase-space transformation). These results were recently confirmed analytically by Brizard and Hodgeman \cite{Brizard_Hodgeman_2023} for general axisymmetric magnetic geometry.

\section{\label{sec:gc_var}Variational formulation of the guiding-center Vlasov-Maxwell equations}

In this Section, we present the variational formulation of the guiding-center Vlasov-Maxwell equations based on the Eulerian guiding-center action functional \cite{Newcomb_1962,Grad_1966,Scott_2010,Tronko_2018,McMillan_2023}
\begin{equation}
{\cal A}_{\rm gc} = -\,\int {\cal F}_{\rm gc}{\cal H}_{\rm gc}\,d^{8}Z - \int \frac{|{\bf B}|^{2}}{8\pi}\;d^{4}X,
\label{eq:A_gc}
\end{equation}
where the electric-field Lagrangian density $|{\bf E}|^{2}/8\pi$ has been removed \cite{Brizard_2000_prl}, which explicitly imposes quasineutrality and eliminates the displacement current density $c^{-1}\partial{\bf E}/\partial t$ from the guiding-center Maxwell equations, while the negative signs in Eq.~\eqref{eq:A_gc} are consistent with the energy-momentum conservation laws. Here, the extended guiding-center Vlasov phase-space density 
\begin{equation}
{\cal F}_{\rm gc} \;\equiv\; {\cal J}_{\rm gc}\,F\;\delta(W - H_{\rm gc})
\label{eq:Fgc_def}
\end{equation}
includes the guiding-center Vlasov function $F$ and the guiding-center Jacobian ${\cal J}_{\rm gc}$, while the delta-function ensures that the extended guiding-center Hamiltonian motion takes place on the energy surface ${\cal H}_{\rm gc} = H_{\rm gc} - W \equiv 0$. This expression leads to the integral constraint
\begin{equation}
\int {\cal F}_{\rm gc}\;{\cal H}_{\rm gc}\;{\cal G}\;d^{8}Z \;\equiv\; 0,
\label{eq:FHG_id}
\end{equation}
where ${\cal G}$ is an arbitrary extended phase-space function.

The variation of the  guiding-center action functional \eqref{eq:A_gc} is expressed as
\begin{eqnarray}
\delta{\cal A}_{\rm gc} &=& -\;\int \left(\delta{\cal F}_{\rm gc}\;{\cal H}_{\rm gc} \;+\frac{}{} {\cal F}_{\rm gc}\;\delta{\cal H}_{\rm gc}\right)d^{8}Z \nonumber \\
 &&-\; \int \frac{{\bf B}}{4\pi}\bdot\delta{\bf B}\;d^{4}X,
 \label{eq:delta_Agc}
 \end{eqnarray}
 where the Eulerian variation of the extended guiding-center Vlasov phase-space density is expressed as
 \begin{equation}
 \delta{\cal F}_{\rm gc} \;\equiv\; -\;\pd{}{Z^{\alpha}}\left({\cal F}_{\rm gc}\;\delta Z^{\alpha}\right),
   \label{eq:delta_Fgc}
  \end{equation}
  where the virtual phase-space displacement
 \begin{equation}
 \delta Z^{\alpha} \;\equiv\; \left\{ Z^{\alpha},\frac{}{} \delta{\sf S}\right\}_{\rm gc} \;-\; \frac{e}{c}\,\delta{\bf A}^{*}\bdot\left\{ {\bf X},\;Z^{\alpha}\right\}_{\rm gc}
 \end{equation}
 is generated by the virtual canonical generating function $\delta{\sf S}$, while the Eulerian variation of the extended guiding-center Hamiltonian is expressed as
\begin{equation}
\delta{\cal H}_{\rm gc} \;=\; e\,\delta\Phi \;+\; \delta K_{\rm gc} \;+\; J\,\delta{\cal S},
\label{eq:delta_Ham}
\end{equation}
while
\begin{eqnarray}
\frac{e}{c}\delta{\bf A}^{*} &=& \frac{e}{c}\delta{\bf A} \;+\; \delta\vb{\Pi}_{\rm gc} \;-\; J\,\delta{\bf R}. \label{eq:deltaA_star}
 \end{eqnarray}
Here, the variations $\delta K_{\rm gc}$ and $\delta\vb{\Pi}_{\rm gc}$ are expressed as
\begin{eqnarray}
\delta K_{\rm gc} &=& \mu\,\bhat\bdot\delta{\bf B} \;+\; {\bf u}_{\rm E}\bdot\delta\vb{\Pi}_{\rm E} \nonumber \\
 &&-\; \nabla\bdot\left[ \frac{J}{2} \left(\delta\bhat\btimes{\bf u}_{\rm E} + \bhat\btimes\delta{\bf u}_{\rm E}\right)\right], \label{eq:delta_K} \\
\delta\vb{\Pi}_{\rm gc} &=& P_{\|}\,\delta\bhat \;+\; \delta\vb{\Pi}_{\rm E} \;-\; \nabla\btimes\left(\frac{J}{2}\,\delta\bhat\right), \label{eq:delta_Pi}
\end{eqnarray}
 which contain the Eulerian variations
 \begin{eqnarray}
 \delta\vb{\Pi}_{\rm E} &=& \delta{\bf E}\btimes(e\bhat/\Omega) \;+\; mc\,{\bf E}\btimes\delta\left(\bhat/B\right) \nonumber \\
  &=& \delta{\bf E}\btimes(e\bhat/\Omega) - \left(\bhat\,\vb{\Pi}_{\rm E} + \vb{\Pi}_{\rm E}\,\bhat\right)\bdot\delta{\bf B}/B,
 \end{eqnarray}
and
 \begin{equation}
 \delta\bhat \;=\; \delta({\bf B}/B) \;\equiv\; \left(\mathbb{I} - \bhat\bhat\right)\bdot\delta{\bf B}/B, 
 \end{equation}
while the gyrogauge-field variations
 \begin{equation}
 \left(\delta{\cal S}, \delta{\bf R}\right) \;=\; \left(-\,\pd{\bhat}{t}\btimes\frac{\bhat}{B}\bdot\delta{\bf B},\; -\,\nabla\bhat\btimes\frac{\bhat}{B}\bdot\delta{\bf B}\right)
 \end{equation}
 are included in Eqs.~\eqref{eq:delta_Ham}-\eqref{eq:deltaA_star}.
 
When Eqs.~\eqref{eq:delta_Fgc} and \eqref{eq:delta_Ham} are inserted in Eq.~\eqref{eq:delta_Agc}, we obtain
 \begin{eqnarray}
 \delta\left({\cal F}_{\rm gc}\;{\cal H}_{\rm gc} \right) &=& -\;{\cal F}_{\rm gc}\,\delta L_{\rm gc} + B_{\|}^{*}\,\delta{\sf S} \left\{ {\cal F}_{\rm gc}/B_{\|}^{*},\frac{}{} {\cal H}_{\rm gc}\right\}_{\rm gc} \nonumber \\
  &&-\; \pd{}{Z^{\alpha}}\left({\cal F}_{\rm gc}\,\delta{\sf S}\;\left\{ Z^{\alpha},\frac{}{} {\cal H}_{\rm gc}\right\}_{\rm gc}\right),
  \label{eq:delta_FH}
  \end{eqnarray}
 where the variation of the guiding-center Lagrangian \eqref{eq:Lag_gc_def} is expressed as
 \begin{eqnarray}
\delta L_{\rm gc} &\equiv& \left( \frac{e}{c}\delta{\bf A} \;+\; \delta\vb{\Pi}_{\rm gc} \right)\bdot\dot{\bf X} \;-\; \left(e\,\delta\Phi \;+\frac{}{} \delta K_{\rm gc} \right) \nonumber \\
  &&-\; J \left(\delta{\cal S} \;+\frac{}{} \dot{\bf X}\bdot\delta{\bf R}\right), 
  \label{eq:delta_Psi}
  \end{eqnarray}
with the effective gyrogauge variation
\begin{eqnarray} 
\delta{\cal S} + \dot{\bf X}\bdot\delta{\bf R} &=& -\;\left(\pd{\bhat}{t} + \dot{\bf X}\bdot\nabla\bhat\right)\btimes\frac{\bhat}{B}\bdot\delta{\bf B} \\
 &=& -\;\frac{d\bhat}{dt}\btimes\frac{\bhat}{B}\bdot\delta{\bf B}.
 \label{eq:effective_gyro}
 \end{eqnarray}
Finally, when Eq.~\eqref{eq:delta_FH} is inserted in Eq.~\eqref{eq:delta_Agc}, the variation of the guiding-center action functional may be expressed as $\delta{\cal A}_{\rm gc} \equiv \int \delta{\cal L}_{\rm gc}\,d^{3}X dt$, where the variation of the guiding-center Lagrangian density is expressed as
  \begin{eqnarray}
\delta{\cal L}_{\rm gc} &=& \int \left({\cal F}_{\rm gc}\,\delta L_{\rm gc} - B_{\|}^{*}\,\delta{\sf S} \left\{ {\cal F}_{\rm gc}/B_{\|}^{*},\frac{}{} {\cal H}_{\rm gc}\right\}_{\rm gc} \right)d^{4}P \nonumber \\
 &&-\; \frac{{\bf B}}{4\pi}\bdot\delta{\bf B} \;+\; \delta\Lambda_{\rm gcV},
 \label{eq:delta_Lgc_eq}
 \end{eqnarray}
 where $d^{4}P \equiv 2\pi\,dP_{\|}dJ\,dW$ (which excludes the guiding-center Jacobian ${\cal J}_{\rm gc}$), the guiding-center position ${\bf X}$ is now the location where the electromagnetic fields ${\bf E}$ and ${\bf B}$ are evaluated, and the last term in Eq.~\eqref{eq:delta_Lgc_eq} represents the space-time derivative of the guiding-center Vlasov action-density variation
 \begin{equation}
\delta\Lambda_{\rm gcV} \equiv \pd{}{t}\left(\int {\cal F}_{\rm gc}\,\delta{\sf S}\,d^{4}P\right) + \nabla\vb{\cdot}\left( \int {\cal F}_{\rm gc}\,\delta{\sf S}\;\dot{\bf X}\;d^{4}P\right)
 \end{equation}
 generated by $\delta{\sf S}$.
   
  \subsection{Guiding-center Vlasov equation}
 
 Variation of the guiding-center action functional $\delta{\cal A}_{\rm gc}$ with respect to $\delta{\sf S}$ yields the extended guiding-center Vlasov equation
 \begin{eqnarray}
 0 &=& B_{\|}^{*}\left\{ {\cal F}_{\rm gc}/B_{\|}^{*},\frac{}{} {\cal H}_{\rm gc}\right\}_{\rm gc}  \nonumber \\
  &=& \pd{}{Z^{\alpha}}\left( {\cal F}_{\rm gc}\;\left\{ Z^{\alpha},\frac{}{} {\cal H}_{\rm gc} \right\}_{\rm gc}\right).
 \end{eqnarray}
 When this extended Vlasov equation is integrated over the guiding-center energy coordinate $W$, using Eq.~\eqref{eq:Fgc_def}, we obtain the phase-space divergence form of the guiding-center Vlasov equation
  \begin{equation}
 \pd{F_{\rm gc}}{t} \;+\; \nabla\bdot\left(F_{\rm gc}\frac{}{}\dot{\bf X}\right) \;+\; \pd{}{P_{\|}}\left(F_{\rm gc}\frac{}{}\dot{P}_{\|}\right) \;=\; 0,
 \label{eq:gcVlasov_div}
 \end{equation}
which yields the guiding-center Vlasov equation 
\begin{equation}
\pd{F}{t} \;+\; \dot{\bf X}\bdot\nabla F \;+\; \dot{P}_{\|}\;\pd{F}{P_{\|}} \;=\; 0, 
\label{eq:gcVlasov}
\end{equation}
when we substitute the definition of the guiding-center Vlasov density $F_{\rm gc} \equiv {\cal J}_{\rm gc}F$ and make use of the guiding-center Liouville equation \eqref{eq:gc_Liouville}.
  
  \subsection{\label{sec:Max_gc}Guiding-center Maxwell equations}
  
 Variations of the guiding-center action functional $\delta{\cal A}_{\rm gc}$ with respect to $\delta\Phi$ and $\delta{\bf A}$ in Eq.~\eqref{eq:delta_Psi}, respectively, yield the guiding-center charge and current densities
  \begin{equation}
  (\varrho_{\rm gc}, {\bf J}_{\rm gc}) \;=\; \int_{P} F_{\rm gc} \left(e,\frac{}{} e\,\dot{\bf X}\right),
  \label{eq:rho_J_gc}
  \end{equation}
where we use the notation $\int_{P} (\cdots) \equiv \int (\cdots)\; 2\pi\,dP_{\|}dJ$.  We note that, in Eq.~\eqref{eq:delta_Psi}, the guiding-center Lagrangian \eqref{eq:Lag_gc_def} also depends on the electric and magnetic fields $({\bf E},{\bf B})$, which introduces guiding-center polarization and magnetization charge and current densities in the guiding-center Maxwell equations \cite{Brizard_2008}.

\subsubsection{Guiding-center polarization}

When a virtual electric variation $\delta_{\bf E}L_{\rm gc}$ of the guiding-center Lagrangian \eqref{eq:Lag_gc_def} associated with $\delta{\bf E}$ is considered in Eq.~\eqref{eq:delta_Psi}, we find
\begin{eqnarray}
\delta_{\bf E}L_{\rm gc}  &=& \vb{\pi}_{\rm gc}\bdot\delta{\bf E} \;+\; \nabla\bdot\left( \mathbb{Q}_{\rm gc}\bdot\delta{\bf E}\right),
\end{eqnarray}
where the guiding-center electric dipole moment \cite{Pfirsch_1984,Pfirsch_Morrison_1985}
\begin{equation}
\vb{\pi}_{\rm gc} \;\equiv\; \frac{e\bhat}{\Omega}\btimes\left(\dot{\bf X} \;-\frac{}{} {\bf u}_{\rm E}\right),
\label{eq:pi_gc}
\end{equation}
which, when Eq.~\eqref{eq:xdot_P0} is substituted, is approximated as
\begin{equation}
\vb{\pi}_{\rm gc} \;=\; -\,\frac{c}{B_{\|}^{*}\Omega} \left(\frac{d_{0}{\bf P}_{0}}{dt} \;+\; \mu\;\nabla B\right)_{\bot},
\label{eq:pi0_gc}
\end{equation}
and the guiding-center electric quadrupole moment is defined by the symmetric dyadic tensor
\begin{equation}
\mathbb{Q}_{\rm gc} \;\equiv\; \frac{e}{2}\,\langle\vb{\rho}_{0}\vb{\rho}_{0}\rangle \;=\; \frac{J\,c}{2B}\;\left(\mathbb{I} \;-\; \bhat\,\bhat\right).
\label{eq:Qgc_def}
\end{equation}
We can thus define the guiding-center polarization as
\begin{equation}
\vb{\cal P}_{\rm gc} \;\equiv\; \int_{P} \left( F_{\rm gc}\,\vb{\pi}_{\rm gc} \;-\; \mathbb{Q}_{\rm gc}\bdot\nabla F_{\rm gc}\right),
\label{eq:Pol_gc}
\end{equation}
which includes a finite-Larmor-radius (FLR) correction on the guiding-center Vlasov phase-space density $F_{\rm gc}$. Hence, in Eq.~\eqref{eq:delta_Lgc_eq}, we find
\begin{equation}
\int_{P}F_{\rm gc}\;\delta_{\bf E}L_{\rm gc}  \;=\; \vb{\cal P}_{\rm gc}\bdot\delta{\bf E} \;+\; \delta\Lambda_{\rm gcP},
\label{eq:Psi_E}
\end{equation}
with the polarization spatial divergence defined as
\begin{equation}
\delta\Lambda_{\rm gcP} \;\equiv\; \nabla\bdot\left(\int_{P}F_{\rm gc}\;\mathbb{Q}_{\rm gc}\bdot\delta{\bf E}\right).
\end{equation}

\subsubsection{Guiding-center magnetization}

When a virtual magnetic variation $\delta_{\bf B}L_{\rm gc}$ of the guiding-center Lagrangian \eqref{eq:Lag_gc_def} associated with $\delta{\bf B}$ is considered in Eq.~\eqref{eq:delta_Psi}, we find
\begin{eqnarray}
\delta_{\bf B}L_{\rm gc}  &=& \left( \vb{\mu}_{\rm gc} \;+\; \vb{\pi}_{\rm gc}\btimes\frac{{\bf P}_{0}}{mc}\right)\bdot\delta{\bf B} \nonumber \\
 &&+\; \nabla\bdot\left[ \mathbb{Q}_{\rm gc}\bdot\frac{{\bf u}_{\rm E}}{c}\btimes\delta{\bf B} \;+\; \left(\mathbb{Q}_{\rm gc}\bdot\delta{\bf B}\right)\btimes\frac{{\bf u}_{\rm E}}{c}\right] \nonumber \\
  &&-\; \frac{1}{c}\dot{\bf X}\bdot\nabla\btimes\left(\mathbb{Q}_{\rm gc}\bdot\delta{\bf B}\right),
\end{eqnarray}
 where  the intrinsic magnetic dipole moment
\begin{equation}
\vb{\mu}_{\rm gc} \;\equiv\; \mu \left(-\,\bhat \;+\; \frac{1}{\Omega}\,\frac{d\bhat}{dt}\btimes\bhat\right)
\label{eq:mu_gc}
\end{equation}
includes the effective gyrogauge contribution \eqref{eq:effective_gyro}, and the moving electric-dipole contribution \cite{Jackson_1975}
\begin{equation} 
\vb{\pi}_{\rm gc}\btimes\frac{{\bf P}_{0}}{mc} \;=\; \left[\frac{\bhat}{B}\btimes\left(\dot{\bf X} - {\bf u}_{\rm E}\right)\right]\btimes{\bf P}_{0}
\label{eq:pi_gc_mag}
\end{equation}
is defined in terms of the guiding-center electric dipole moment \eqref{eq:pi_gc}, which can be approximated with Eq.~\eqref{eq:pi0_gc}. We can thus define the guiding-center magnetization as
\begin{eqnarray}
\vb{\cal M}_{\rm gc} &\equiv& \int_{P} F_{\rm gc}\left( \vb{\mu}_{\rm gc} \;+\frac{}{} \vb{\pi}_{\rm gc}\btimes{\bf P}_{0}/mc\right) \nonumber \\
 &&-\; \int_{P} \nabla F_{\rm gc}\bdot\left( \mathbb{Q}_{\rm gc}\btimes\frac{{\bf u}_{\rm E}}{c} \;-\; \frac{{\bf u}_{\rm E}}{c}\btimes\mathbb{Q}_{\rm gc} \right) \nonumber \\
  &&-\; \int_{P} \mathbb{Q}_{\rm gc}\bdot\nabla\btimes\left(F_{\rm gc}\;\dot{\bf X}/c\right),
  \label{eq:Mag_gc}
\end{eqnarray}
which also includes FLR corrections. Hence, in Eq.~\eqref{eq:delta_Lgc_eq}, we find
\begin{eqnarray}
\int_{P}F_{\rm gc}\;\delta_{\bf B}L_{\rm gc} &=& \vb{\cal M}_{\rm gc}\bdot\delta{\bf B} \;+\; \delta\Lambda_{\rm gcM},
\label{eq:Psi_B}
\end{eqnarray}
where the magnetization spatial divergence $\delta\Lambda_{\rm gcM}$ is defined as
\begin{eqnarray}
\delta\Lambda_{\rm gcM} &\equiv& \nabla\bdot\left[\int_{P} F_{\rm gc}\;\left(\mathbb{Q}_{\rm gc}\btimes\frac{{\bf u}_{\rm E}}{c}\right)\bdot\delta{\bf B} \right. \nonumber \\
 &&\left.- \int_{P} F_{\rm gc}\left(\mathbb{Q}_{\rm gc}\vb{\cdot}\delta{\bf B}\right)\vb{\times}\frac{1}{c}(\dot{\bf X} - {\bf u}_{\rm E}) \right].
\end{eqnarray}

\subsubsection{Guiding-center Maxwell equations}

By substituting the variations \eqref{eq:Psi_E} and \eqref{eq:Psi_B} into Eq.~\eqref{eq:delta_Lgc_eq}, we find
\begin{eqnarray}
\delta{\cal L}_{\rm gc} &=& \delta\Lambda_{\rm gcV} \;+\; \delta\Lambda_{\rm gcPM} \;-\; \left(\varrho_{\rm gc}\;\delta\Phi \;-\; \frac{{\bf J}_{\rm gc}}{c}\bdot\delta{\bf A}\right) \nonumber \\
 &&+\; \vb{\cal P}_{\rm gc}\bdot\delta{\bf E} \;-\; \frac{{\bf H}_{\rm gc}}{4\pi}\bdot\delta{\bf B},
 \label{eq:delta_Lgc_polmag}
 \end{eqnarray}
 where the guiding-center displacement field ${\bf D}_{\rm gc}$ is replaced with $4\pi\,\vb{\cal P}_{\rm gc}$, which follows from the removal of the electric-field Lagrangian density in Eq.~\eqref{eq:A_gc}, and the guiding-center magnetic field is defined as
 \begin{equation}
 {\bf H}_{\rm gc} \;\equiv\; {\bf B} \;-\; 4\pi\,\vb{\cal M}_{\rm gc}, \label{eq:H_gc} 
\end{equation}
while the quadrupole contributions yield an additional polarization-magnetization spatial divergence
\begin{eqnarray}
\delta\Lambda_{\rm gcPM} &=& \nabla\bdot\left[\int_{P} F_{\rm gc} \;\mathbb{Q}_{\rm gc}\bdot\left(\delta{\bf E} + \frac{{\bf u}_{\rm E}}{c}\btimes\delta{\bf B}\right) \right. \nonumber \\
 &&\left.- \int_{P} F_{\rm gc} \left(\mathbb{Q}_{\rm gc}\bdot\delta{\bf B}\right)\btimes\frac{1}{c}\left(\dot{\bf X} - {\bf u}_{\rm E}\right)\right].
\end{eqnarray}

If we now introduce the constrained variations $\delta{\bf E} = -\,\nabla\delta\Phi - c^{-1}\partial\delta{\bf A}/\partial t$ and $\delta{\bf B} = \nabla\btimes\delta{\bf A}$ into Eq.~\eqref{eq:delta_Lgc_polmag}, we obtain the guiding-center Lagrangian variation
\begin{eqnarray}
\delta{\cal L}_{\rm gc} &=& \delta\Lambda_{\rm gc} \;+\; \delta\Phi \left( \nabla\bdot\vb{\cal P}_{\rm gc} \;-\frac{}{} \varrho_{\rm gc} \right) \nonumber \\
 &&+\; \frac{1}{c}\,\delta{\bf A}\bdot\left( {\bf J}_{\rm gc} + \pd{\vb{\cal P}_{\rm gc}}{t} - \frac{c}{4\pi}\;\nabla\btimes{\bf H}_{\rm gc} \right),
\label{eq:deltaL_gc_Maxwell} 
\end{eqnarray}
where the total guiding-center space-time divergence is
\begin{eqnarray}
\delta\Lambda_{\rm gc} &\equiv& \pd{}{t} \left( \int {\cal F}_{\rm gc}\,\delta{\sf S}\;d^{4}P \;-\; \frac{1}{c}\,\delta{\bf A}\bdot\vb{\cal P}_{\rm gc} \right)  \\
 &&+ \nabla\vb{\cdot}\left( \int {\cal F}_{\rm gc}\delta{\sf S}\,\dot{\bf X}\;d^{4}P - \delta\Phi\,\vb{\cal P}_{\rm gc} - \frac{\delta{\bf A}}{4\pi}\vb{\times}{\bf H}_{\rm gc} \right) \nonumber \\
  &&+ \nabla\bdot\left[\int_{P} F_{\rm gc} \;\mathbb{Q}_{\rm gc}\bdot\left(\delta{\bf E} + \frac{{\bf u}_{\rm E}}{c}\btimes\delta{\bf B}\right) \right. \nonumber \\
 &&\left.\hspace*{0.5in}-\; \int_{P} F_{\rm gc} \left(\mathbb{Q}_{\rm gc}\bdot\delta{\bf B}\right)\btimes\frac{1}{c}\left(\dot{\bf X} - {\bf u}_{\rm E}\right)\right]. \nonumber
 \end{eqnarray}
Variations of the guiding-center action functional $\delta{\cal A}_{\rm gc} = \int \delta{\cal L}_{\rm gc}\,d^{3}X\,dt$ with respect to $\delta\Phi$ and $\delta{\bf A}$ yield, respectively, the guiding-center quasineutrality condition
\begin{equation}
0 \;=\; \varrho_{\rm gc} \;-\; \nabla\bdot\vb{\cal P}_{\rm gc}, \label{eq:gc_QN}
\end{equation}
and the guiding-center Maxwell equation
\begin{equation}
\nabla\btimes{\bf H}_{\rm gc} \;=\; \frac{4\pi}{c}\;\left({\bf J}_{\rm gc} + \pd{\vb{\cal P}_{\rm gc}}{t}\right), \label{eq:gc_Maxwell}
\end{equation} 
which excludes the displacement current density $c^{-1}\partial{\bf E}/\partial t$ following the removal of the electric-field Lagrangian density in Eq.~\eqref{eq:A_gc}. In addition, the electromagnetic fields $({\bf E},{\bf B})$ satisfy Faraday's law
\begin{equation}
\partial{\bf B}/\partial t \;=\; -\,c\;\nabla\btimes{\bf E},
\label{eq:Faraday}
\end{equation}
and Gauss's law $\nabla\bdot{\bf B} = 0$. Finally, the guiding-center charge-current densities \eqref{eq:rho_J_gc} satisfy the charge conservation law 
\begin{equation}
\pd{\varrho_{\rm gc}}{t} \;+\; \nabla\bdot{\bf J}_{\rm gc} = 0,
\label{eq:charge_cons}
\end{equation}
which can be obtained directly from Eqs.~\eqref{eq:gc_QN}-\eqref{eq:gc_Maxwell}.

\section{\label{sec:gc_Noether}Guiding-center Noether Equation and Exact Conservation Laws}

Once the guiding-center Vlasov-Maxwell equations \eqref{eq:gcVlasov_div} and \eqref{eq:gc_QN}-\eqref{eq:gc_Maxwell} are derived from the action functional \eqref{eq:A_gc}, we are left with the guiding-center Noether equation
\begin{eqnarray}
\delta{\cal L}_{\rm gc} &=& \pd{}{t} \left( \int {\cal F}_{\rm gc}\,\delta{\sf S}\;d^{4}P \;-\; \frac{1}{c}\,\delta{\bf A}\bdot\vb{\cal P}_{\rm gc} \right) \label{eq:gc_Noether} \\
 &&+ \nabla\vb{\cdot}\left( \int {\cal F}_{\rm gc}\delta{\sf S}\,\dot{\bf X}\;d^{4}P - \delta\Phi\,\vb{\cal P}_{\rm gc} - \frac{\delta{\bf A}}{4\pi}\vb{\times}{\bf H}_{\rm gc}\right) \nonumber \\
   &&+ \nabla\bdot\left[\int_{P} F_{\rm gc} \;\mathbb{Q}_{\rm gc}\bdot\left(\delta{\bf E} + \frac{{\bf u}_{\rm E}}{c}\btimes\delta{\bf B}\right) \right. \nonumber \\
 &&\left.-\; \int_{P} F_{\rm gc} \left(\mathbb{Q}_{\rm gc}\bdot\delta{\bf B}\right)\btimes\frac{1}{c}\left(\dot{\bf X} - {\bf u}_{\rm E}\right)\right], \nonumber
 \end{eqnarray}
from which exact energy-momentum conservation laws are derived. 

For this purpose, we use the expressions
\begin{eqnarray}
\delta{\sf S} &=& (e/c)\,{\bf A}^{*}\bdot\delta{\bf X} \;+\; J\,\delta\Theta \;-\; W^{*}\;\delta t, \\
\delta\Phi &=& {\bf E}\bdot\delta{\bf X} \;-\; c^{-1}\partial\delta\varphi/\partial t, \\
\delta{\bf A} &=& {\bf E}\; c\,\delta t \;+\; \delta{\bf X}\btimes{\bf B} \;+\; \nabla\delta\varphi,
\end{eqnarray}
where $\delta{\sf S}$ is gyrogauge-independent, with the virtual gauge variations $\delta\Theta \equiv {\bf R}\bdot\delta{\bf X} + {\cal S}\,\delta t$ and $\delta\varphi \equiv \Phi\;c\,\delta t - {\bf A}\bdot\delta{\bf X}$. The gauge terms associated with $\delta\varphi$ can be used to obtain the identity
\begin{eqnarray}
 &&-\;\pd{}{t} \left(\nabla\delta\varphi\bdot\vb{\cal P}_{\rm gc} \right) + \nabla\vb{\cdot}\left(\pd{\delta\varphi}{t}\,\vb{\cal P}_{\rm gc} - \nabla\delta\varphi\vb{\times} \frac{c{\bf H}_{\rm gc}}{4\pi}\right) \nonumber \\
   &&\hspace{0.2in}\equiv\; \pd{}{t}\left(\delta\varphi\frac{}{}\varrho_{\rm gc}\right) \;+\; \nabla\bdot\left(\delta\varphi\frac{}{}{\bf J}_{\rm gc}\right),
   \label{eq:em_gauge_id}
\end{eqnarray}
where we have used the guiding-center Maxwell equations  \eqref{eq:gc_QN}-\eqref{eq:gc_Maxwell} in order to obtain the last equality.

\subsection{Guiding-center Noether equation}

By substituting the gauge identity \eqref{eq:em_gauge_id} into Eq.~\eqref{eq:gc_Noether}, the guiding-center Noether equation becomes
\begin{equation}
\delta{\cal L}_{\rm gc} \;\equiv\; \partial\delta{\cal N}_{\rm gc}/\partial t \;+\; \nabla\bdot\delta{\bf F}_{\rm gc},
\label{eq:gc_Noether_final}
\end{equation}
where the guiding-center action-density variation is defined as
\begin{eqnarray}
\delta{\cal N}_{\rm gc} &\equiv& \int_{P} F_{\rm gc} \left(\delta{\sf S} \;+\; \frac{e}{c}\,\delta\varphi\right) \nonumber \\
 &&-\; \frac{1}{c}\,\left( {\bf E}\;c\,\delta t \;+\frac{}{} \delta{\bf X}\btimes{\bf B}\right)\bdot\vb{\cal P}_{\rm gc}
\end{eqnarray}
and the guiding-center action-density-flux variation is defined as
\begin{eqnarray}
\delta{\bf F}_{\rm gc} &\equiv& \int_{P} F_{\rm gc} \left(\delta{\sf S} \;+\; \frac{e}{c}\,\delta\varphi\right)\,\dot{\bf X} \;-\; \delta{\bf X}\bdot\left({\bf E}\frac{}{}\vb{\cal P}_{\rm gc}\right) \nonumber \\
 &&-\; \left( {\bf E}\;c\,\delta t \;+\frac{}{} \delta{\bf X}\btimes{\bf B}\right)\btimes\frac{{\bf H}_{\rm gc}}{4\pi} \nonumber \\
 &&+\; \int_{P} F_{\rm gc} \;\mathbb{Q}_{\rm gc}\bdot\left(\delta{\bf E} + \frac{{\bf u}_{\rm E}}{c}\btimes\delta{\bf B}\right)\nonumber \\
 &&-\; \int_{P} F_{\rm gc} \left(\mathbb{Q}_{\rm gc}\bdot\delta{\bf B}\right)\btimes\frac{1}{c}\left(\dot{\bf X} - {\bf u}_{\rm E}\right).
\end{eqnarray}
Here, the electric and magnetic variations are
\begin{equation}
\left( \begin{array}{c}
\delta{\bf E} \\
\\
\delta{\bf B} \end{array} \right) \;=\;  \left( \begin{array}{c}
-\,\delta t\;\partial{\bf E}/\partial t \;-\; \delta{\bf X}\bdot\nabla{\bf E} \\
\\
 -\,\delta t\;\partial{\bf B}/\partial t \;-\; \delta{\bf X}\bdot\nabla{\bf B} \end{array} \right),
 \label{eq:delta_EB}
\end{equation}
while the combination
\begin{equation}
\delta{\sf S} \;+\; \frac{e}{c}\,\delta\varphi \;\equiv\; \vb{\Pi}_{\rm gc}\bdot\delta{\bf X} \;-\; K_{\rm gc}\;\delta t
\end{equation}
is gauge invariant, where the guiding-center momentum ${\bf P}_{0}$ and the guiding-center kinetic energy $K_{\rm gc}$ are defined in Eqs.~\eqref{eq:Pi_gc_def} and \eqref{eq:K_gc_def}, respectively. Finally, the constrained Lagrangian variation is expressed as
\begin{equation}
\delta{\cal L}_{\rm gc} \;\equiv\; \delta t\,\pd{}{t}\left(\frac{|{\bf B}|^{2}}{8\pi}\right) \;+\; \delta{\bf X}\bdot\nabla\left(\frac{|{\bf B}|^{2}}{8\pi}\right),
\label{eq:delta_LM}
\end{equation}
while the Vlasov contribution in Eq.~\eqref{eq:A_gc} vanishes because of the constraint \eqref{eq:FHG_id}.

\subsection{Exact guiding-center energy-momentum conservation laws}

By considering a virtual time displacement $\delta t$ in the guiding-center Noether equation \eqref{eq:gc_Noether_final}, we obtain the guiding-center Vlasov-Maxwell energy conservation law 
\begin{equation}
\partial{\cal E}_{\rm gc}/\partial t \;+\; \nabla\bdot{\bf S}_{\rm gc} \;=\; 0,
\label{eq:energy_law}
\end{equation}
which is expressed in terms of the guiding-center Vlasov-Maxwell energy density ${\cal E}_{\rm gc} \equiv -\;\delta{\cal N}_{\rm gc}/\delta t - {\cal L}_{\rm M}$ and the guiding-center Vlasov-Maxwell energy density-flux ${\bf S}_{\rm gc} \equiv -\,
\delta{\bf F}_{\rm gc}/\delta t$, where
\begin{eqnarray}
{\cal E}_{\rm gc} &=& \int_{P} F_{\rm gc}\,K_{\rm gc} + {\bf E}\bdot\vb{\cal P}_{\rm gc} \;+\; |{\bf B}|^{2}/8\pi,
 \label{eq:Egc_def}
\end{eqnarray}
and
\begin{eqnarray}
{\bf S}_{\rm gc} &=& \int_{P} F_{\rm gc}\,K_{\rm gc}\,\dot{\bf X} \;+\; \frac{c\,{\bf E}}{4\pi}\btimes{\bf H}_{\rm gc} 
\label{eq:Sgc_def} \\
&&+\; \int_{P} F_{\rm gc} \;\mathbb{Q}_{\rm gc}\bdot\left(\pd{\bf E}{t} \;+\; \frac{{\bf u}_{\rm E}}{c}\btimes\pd{\bf B}{t} \right) \nonumber \\
 &&-\; \int_{P} F_{\rm gc} \left(\mathbb{Q}_{\rm gc}\bdot\pd{\bf B}{t}\right)\btimes\frac{1}{c}\left(\dot{\bf X} - {\bf u}_{\rm E}\right), \nonumber
\end{eqnarray}
which includes FLR corrections involving $\mathbb{Q}_{\rm gc}$. We note that, in the case of static electric and magnetic fields, only the terms on the first line on the right side of Eq.~\eqref{eq:Sgc_def} remain. 

Next, by considering a virtual spatial displacement $\delta X^{k}$ in the guiding-center Noether equation \eqref{eq:gc_Noether_final}, we obtain the guiding-center Vlasov-Maxwell momentum conservation law 
\begin{equation}
\partial P_{{\rm gc}k}/\partial t \;+\; \nabla\bdot{\bf T}_{{\rm gc}k} \;=\; 0
\label{eq:momentum_law}
\end{equation}
in the $X^{k}$-direction is expressed in terms of the guiding-center Vlasov-Maxwell momentum density 
\begin{equation}
\vb{\sf P}_{\rm gc} \;=\; \int_{P} F_{\rm gc}\,\vb{\Pi}_{\rm gc} \;+\; \vb{\cal P}_{\rm gc}\btimes \frac{1}{c}\,{\bf B},
\label{eq:Pgck_def}
\end{equation}
where $P_{{\rm gc}k} \equiv \delta{\cal N}_{\rm gc}/\delta X^{k} = \vb{\sf P}_{\rm gc}\bdot\partial_{k}{\bf X}$, and the guiding-center Vlasov-Maxwell momentum-density flux $T_{{\rm gc}k}^{i} \equiv \delta F_{\rm gc}^{i}/\delta X^{k} + \delta^{i}_{k}\;{\cal L}_{\rm M}$, where
 \begin{widetext}
\begin{eqnarray}
{\bf T}_{{\rm gc}k} \;\equiv\; \mathbb{T}_{\rm gc}\bdot\partial_{k}{\bf X} &=& \int_{P} F_{\rm gc}\,\dot{\bf X}\,\Pi_{{\rm gc}k} \;-\; \left(\vb{\cal P}_{\rm gc}\,E_{k} + \frac{\bf B}{4\pi}\,{\sf H}_{{\rm gc}k}\right) \;+\; \frac{1}{4\pi} \partial_{k}{\bf X}\left( {\bf B}\bdot{\bf H}_{\rm gc} \;-\; \frac{1}{2}\,|{\bf B}|^{2} \right) \nonumber \\
 &&-\; \int_{P} F_{\rm gc} \left[\mathbb{Q}_{\rm gc}\bdot\left(\partial_{k}{\bf E} + \frac{{\bf u}_{\rm E}}{c}\btimes\partial_{k}{\bf B}\right) \;-\; \left(\mathbb{Q}_{\rm gc}\bdot\partial_{k}{\bf B}\right)\btimes\frac{1}{c}\left(\dot{\bf X} - {\bf u}_{\rm E}\right)\right],
 \label{eq:Tgck_def}
\end{eqnarray}
\end{widetext}
with the notation $\partial_{k} \equiv (\partial{\bf X}/\partial X^{k})\bdot\nabla$. We note that guiding-center Vlasov-Maxwell stress tensor $\mathbb{T}_{\rm gc}$ defined in Eq.~\eqref{eq:Tgck_def} is manifestly not symmetric, which is quite common in reduced Vlasov-Maxwell systems \cite{Pfirsch_1984,Pfirsch_Morrison_1985}. The symmetry properties of the lowest-order guiding-center Vlasov-Maxwell stress tensor are briefly discussed in Sec.~\ref{sec:sym}. 

The proofs of the energy-momentum conservation laws \eqref{eq:energy_law} and \eqref{eq:momentum_law} proceed by taking the partial time derivative of the guiding-center Vlasov-Maxwell energy density \eqref{eq:Egc_def} and the guiding-center Vlasov-Maxwell momentum density \eqref{eq:Pgck_def} which, after substituting the guiding-center Vlasov-Maxwell equations \eqref{eq:gcVlasov_div} and \eqref{eq:gc_QN}-\eqref{eq:gc_Maxwell}, yield
\begin{widetext}
\begin{eqnarray}
\pd{{\cal E}_{\rm gc}}{t} &=& -\;\nabla\bdot\left(\int_{P} F_{\rm gc}\,K_{\rm gc}\,\dot{\bf X} + \frac{c\,{\bf E}}{4\pi}\btimes{\bf H}_{\rm gc}\right) \;+\; \int_{P} F_{\rm gc} \left(\frac{dK_{\rm gc}}{dt} \;-\; e\,{\bf E}\bdot\dot{\bf X}\right) \;+\; 
\pd{\bf E}{t}\bdot\vb{\cal P}_{\rm gc} \;+\; \pd{\bf B}{t}\bdot\vb{\cal M}_{\rm gc}, \\
  \pd{P_{{\rm gc}k}}{t} &=& -\;\nabla\bdot\left[ \int_{P} F_{\rm gc} \dot{\bf X}\,\Pi_{{\rm gc}k} - \left(\vb{\cal P}_{\rm gc}\,E_{k} + \frac{\bf B}{4\pi}\,{\sf H}_{{\rm gc}k}\right) \right] -\; \partial_{k}\left( \frac{\bf B}{4\pi}\bdot{\bf H}_{\rm gc} - \frac{|{\bf B}|^{2}}{8\pi} \right) \nonumber \\
  &&+\; \int_{P} F_{\rm gc} \left[ \frac{d\Pi_{{\rm gc}k}}{dt} \;-\; \partial_{k}{\bf X}\bdot\left( e\,{\bf E} + \frac{e}{c}\,\dot{\bf X}\btimes{\bf B}\right) \right] \;-\; \left( \partial_{k}{\bf E}\bdot\vb{\cal P}_{\rm gc} \;+\frac{}{} \partial_{k}{\bf B}\bdot\vb{\cal M}_{\rm gc} \right).
\end{eqnarray}
\end{widetext}
The last steps in deriving the energy-momentum conservation laws \eqref{eq:energy_law} and \eqref{eq:momentum_law} involve using the guiding-center Euler-Lagrange equation \eqref{eq:ELgc_X} and substituting the expressions \eqref{eq:Pol_gc} and \eqref{eq:Mag_gc} for the guiding-center polarization and magnetization, respectively. 

Finally, we note that the definitions \eqref{eq:Egc_def}-\eqref{eq:Sgc_def} and \eqref{eq:Pgck_def}-\eqref{eq:Tgck_def} associated with the guiding-center energy and momentum conservation laws \eqref{eq:energy_law} and \eqref{eq:momentum_law}, respectively, are not uniquely defined. Indeed, under the following transformations ${\cal E}_{\rm gc}^{\prime} = {\cal E}_{\rm gc} + \nabla\bdot{\bf C}$ and ${\bf S}_{\rm gc}^{\prime} = {\bf S}_{\rm gc} - \partial{\bf C}/\partial t + \nabla\btimes{\bf K}$, the guiding-center energy conservation law \eqref{eq:energy_law}  remains invariant, where the fields $({\bf C},{\bf K})$ are arbitrary. Likewise, under the transformations $P_{{\rm gc}k}^{\prime} = P_{{\rm gc}k} + \nabla\bdot{\bf G}_{k}$ and ${\bf T}_{{\rm gc}k}^{\prime} = {\bf T}_{{\rm gc}k} - \partial{\bf G}_{k}/\partial t + \nabla\bdot\mathbb{K}_{k}$, the guiding-center momentum conservation law \eqref{eq:momentum_law} remains invariant, where the fields ${\bf G}_{k} \equiv \mathbb{G}\bdot\partial_{k}{\bf X}$ is defined in terms of an arbitrary second-rank tensor $\mathbb{G}$, while the third-rank tensor $\mathbb{K}$ has the following antisymmetry property: ${\sf K}^{ji}_{k} = -\,{\sf K}_{k}^{ij}$ (so that $\partial_{ij}^{2}{\sf K}_{k}^{ij} \equiv 0$). Here, we note that the vector fields ${\bf C} \equiv \int_{P} F_{\rm gc}\,\mathbb{Q}_{\rm gc}\bdot{\bf E}$ and ${\bf G}_{k} \equiv \left(\frac{1}{2} \int_{P} F_{\rm gc}\,J\bhat\right)\btimes\partial_{k}{\bf X} \equiv \mathbb{G}\bdot\partial_{k}{\bf X}$ completely remove the FLR corrections in the definitions \eqref{eq:Egc_def} and \eqref{eq:Pgck_def}.

 \subsection{\label{sec:sym}Symmetry properties of the guiding-center stress tensor}
 
 We now make a few remarks about the symmetry properties of the guiding-center stress tensor \eqref{eq:Tgck_def}. These symmetry properties are most relevant when considering the conservation law of guiding-center toroidal angular momentum, where
the guiding-center toroidal angular momentum density 
\begin{equation}
P_{{\rm gc}\phi} \;\equiv\; \int_{P} F_{\rm gc}\;\vb{\Pi}_{\rm gc}\bdot\pd{\bf X}{\phi}  + \frac{1}{c}\,\vb{\cal P}_{\rm gc}\bdot{\bf B}\btimes\pd{\bf X}{\phi} 
\label{eq:Pgc_phi_def}
\end{equation}
is defined as the covariant component of the guiding-center momentum density $\vb{\sf P}_{\rm gc}$ associated with the toroidal angle $\phi$. 
 
 First, we derive the guiding-center angular momentum transport equation from the guiding-center momentum conservation law
 \begin{eqnarray}
 \pd{P_{{\rm gc}\phi}}{t} \;+\; \nabla\bdot\left(\mathbb{T}_{\rm gc}\bdot\pd{\bf X}{\phi}\right) &=& \mathbb{T}_{\rm gc}^{\top}:\nabla\left(\pd{\bf X}{\phi}\right) \nonumber \\
  &\equiv& \wh{\sf z}\bdot\vb{\cal T}_{\rm gc},
 \label{eq:Pgc_phi}
 \end{eqnarray}
where $\mathbb{T}_{\rm gc}^{\top}$ denotes the transpose of $\mathbb{T}_{\rm gc}$. Since the dyadic tensor $\nabla(\partial{\bf X}/\partial\phi)$ is anti-symmetric, with the rotation axis directed along the ${\sf z}$-axis, the anti-symmetric part of the guiding-center stress tensor \eqref{eq:Tgck_def} generates the guiding-center Vlasov-Maxwell torque $\vb{\cal T}_{\rm gc}$. We note that this equation can also be derived by substituting the virtual displacement $\delta{\bf X} = \delta\phi\;\partial{\bf X}/\partial\phi$ in 
Eq.~\eqref{eq:gc_Noether_final}.

At the lowest order (i.e., keeping terms only up to gyrogauge corrections), the guiding-center Vlasov-Maxwell torque is defined as $\vb{\cal T}_{\rm gc} =  \int_{P} F_{\rm gc}\;\vb{\tau}_{\rm gc}$, where, using the dyadic identity (for two arbitrary vector fields 
${\bf F}$ and ${\bf G}$)
\[ {\bf F}\bdot\nabla(\partial{\bf X}/\partial\phi)\bdot{\bf G} \;=\; \wh{\sf z}\bdot({\bf F}\btimes{\bf G}), \]
we find
\begin{eqnarray}
\vb{\tau}_{\rm gc} &=& \dot{\bf X}\btimes\vb{\Pi}_{\rm gc} \;+\; {\bf E}\btimes\vb{\pi}_{\rm gc} \nonumber \\
 &&+\; {\bf B}\btimes\left(\vb{\mu}_{\rm gc} + \vb{\pi}_{\rm gc}\btimes{\bf P}_{0}/mc\right),
 \label{eq:Ngc_def}
\end{eqnarray}
which is calculated below in the dipole approximation. First, each term is expressed as
\begin{eqnarray*}
\dot{\bf X}\btimes\vb{\Pi}_{\rm gc} &=& \dot{\bf X}\btimes\left( P_{\|}\,\bhat \;+\; {\bf E}\btimes\frac{e\bhat}{\Omega}\right) \\
 &=& P_{\|}\,\left(\dot{\bf X}\btimes\bhat + \frac{c{\bf E}}{B}\right) - \left(\dot{\bf X}\bdot \frac{e{\bf E}}{\Omega}\right)\bhat, \\
 {\bf E}\btimes\vb{\pi}_{\rm gc} &=& {\bf E}\btimes\left[\frac{e\bhat}{\Omega}\btimes\left(\dot{\bf X} - {\bf u}_{\rm E}\right)\right] = \left(\dot{\bf X}\bdot \frac{e{\bf E}}{\Omega}\right)\bhat,
\end{eqnarray*}
and
\begin{eqnarray*}
{\bf B}\btimes\vb{\mu}_{\rm gc} &=& {\bf B}\btimes \mu \left(-\,\bhat + \frac{1}{\Omega}\frac{d\bhat}{dt}\btimes\bhat\right) = J\,\frac{d\bhat}{dt}, \\
{\bf B}\btimes\left(\vb{\pi}_{\rm gc}\btimes\frac{{\bf P}_{0}}{mc}\right) &=& P_{\|}\;\bhat\btimes\left(\dot{\bf X} - {\bf u}_{\rm E}\right) \\
 &=& P_{\|}\,\left(\bhat\btimes\dot{\bf X} \;-\; \frac{c{\bf E}}{B}\right),
\end{eqnarray*}
where we used ${\bf B}\bdot\vb{\pi}_{\rm gc} = 0$. Next, by combining these terms in Eq.~\eqref{eq:Ngc_def}, several cancellations occur and the lowest-order torque density is
\begin{equation}
\vb{\tau}_{\rm gc} \;=\; J\,\frac{d\bhat}{dt} \;=\; J \left( \pd{\bhat}{t} \;+\; \dot{\bf X}\bdot\nabla\bhat\right),
\label{eq:Ngc_dip}
\end{equation}
which recovers a result derived by Ye and Kaufman \cite{Ye_Kaufman_1992}. Hence, substituting Eq.~\eqref{eq:Ngc_dip} into the right side of Eq.~\eqref{eq:Pgc_phi}, we obtain
\begin{equation}
\wh{\sf z}\bdot\vb{\cal T}_{\rm gc} = \pd{}{t}\left(\int_{P} F_{\rm gc}\;J\,b_{z}\right) + \nabla\vb{\cdot}\left(\int_{P} F_{\rm gc}\;\dot{\bf X}\;J\,b_{z}\right),
\label{eq:gc_torque}
\end{equation}
which adds a gyrogauge-independent contribution 
\begin{equation}
-\,J\,b_{z} \;\equiv\; -\,J{\bf R}\bdot\partial{\bf X}/\partial\phi 
\label{eq:J_bz}
\end{equation}
to the guiding-center toroidal angular momentum density \eqref{eq:Pgc_phi_def}. Hence, we note that, in the presence of an axisymmetric magnetic field, we can use the definition \eqref{eq:Pi_gc_def} and the identity ${\bf B}\btimes\partial{\bf X}/\partial\phi 
\equiv -\,\nabla A_{\phi}$, the guiding-center toroidal angular momentum density \eqref{eq:Pgc_phi_def} can then be written as
\begin{eqnarray}
P_{{\rm gc}\phi}  &=& \int_{P} F_{\rm gc} \left( P_{\phi} - \frac{e}{c}\,A_{\phi} + J\,b_{z}\right) \;-\; \frac{1}{c}\,\vb{\cal P}_{\rm gc}\bdot\nabla A_{\phi} \nonumber \\
  &=& \int_{P} F_{\rm gc} \left( P_{\phi} \;+\frac{}{} J\,b_{z}\right) - \nabla\bdot\left( \frac{1}{c}\,A_{\phi}\;\vb{\cal P}_{\rm gc} \right),
\end{eqnarray}
where we used the definition \eqref{eq:P_phi} for the guiding-center canonical toroidal angular momentum and the guiding-center quasineutrality condition \eqref{eq:gc_QN}. By using the guiding-center torque correction \eqref{eq:gc_torque} and the transformation $P_{{\rm gc}\phi}^{\prime} = P_{{\rm gc}\phi} + \nabla\bdot{\bf G}_{\phi}$, where  ${\bf G}_{\phi} = (A_{\phi}/c)\,\vb{\cal P}_{\rm gc}$, we obtain the exact guiding-center toroidal angular momentum conservation law
\begin{eqnarray}
\pd{}{t}\left( \int_{P} F_{\rm gc}\;P_{\phi}\right) &=& -\; \nabla\bdot\left(\int_{P} F_{\rm gc}\;\dot{\bf X}\,P_{\phi}\right) \;+\; \int_{P} F_{\rm gc}\;\dot{P}_{\phi} \nonumber \\
 &=&  -\; \nabla\bdot\left(\int_{P} F_{\rm gc}\;\dot{\bf X}\,P_{\phi}\right),
\end{eqnarray}
which follows from applying the Noether theorem to the guiding-center Euler-Lagrange equation $dP_{\phi}/dt = \partial L_{\rm gc}/\partial\phi \equiv 0$ associated with magnetic axisymmetry.

Finally, returning to the case of a general magnetic field, and omitting the gyrogauge terms and FLR corrections, the guiding-center stress tensor \eqref{eq:Tgck_def} can be expressed in symmetric form as 
\begin{eqnarray}
\mathbb{T}_{\rm gc} &=& \mathbb{P}_{\rm CGL} +\; \int_{P} F_{\rm gc}\,P_{\|} \left( \bhat\,\dot{\bf X}_{\bot} \;+\; \dot{\bf X}_{\bot} \,\bhat\right) 
 \label{eq:Tgc_sym} \\
 &&+\; \left(\frac{|{\bf B}|^{2}}{8\pi}\,\mathbb{I} - \frac{{\bf B}\,{\bf B}}{4\pi}\right) + \chi_{\rm gc}\;({\bf E}\btimes\bhat)\,({\bf E}\btimes\bhat) \nonumber \\
  &&+ \left(\mathbb{I} - \bhat\bhat\right) \vb{\cal P}_{\rm gc}\bdot{\bf E} +  \vb{\cal P}_{\rm gc}\btimes\bhat\;{\bf E}\btimes\bhat - \vb{\cal P}_{\rm gc}\;{\bf E},  
 \nonumber
\end{eqnarray}
where $\mathbb{P}_{\rm CGL}$ denotes the Chew-Goldberger-Low (CGL) pressure tensor \cite{CGL_1956}
\[ \mathbb{P}_{\rm CGL} \;=\; \int_{P} F_{\rm gc} \left[ \frac{P_{\|}^{2}}{m}\,\bhat\bhat + \left(\mathbb{I} - \bhat\bhat\right)\;\mu B \right], \]
which appears naturally in lowest-order guiding-center Vlasov-Maxwell theory \citep{Cary_Brizard_2009} and several kinetic-magnetohydrodynamic models \cite{Frieman_1966,Kulsrud_1983,Brizard_2018}, while $\chi_{\rm gc} \equiv \int_{P} F_{\rm gc} (mc^{2}/B^{2})$ denotes the guiding-center electric susceptibility \cite{Northrop1963}. The remaining terms in Eq.~\eqref{eq:Tgc_sym} involve the off-diagonal terms $P_{\|}( \bhat\,\dot{\bf X}_{\bot} + \dot{\bf X}_{\bot} \,\bhat)$, which have appeared in previous works \cite{Sugama_2016,Brizard_Tronci_2016}, and contributions from the guiding-center polarization \eqref{eq:Pol_gc}.  Future work will explore the symmetry properties of the remaining higher-order terms in the guiding-center stress tensor defined in Eq.~\eqref{eq:Tgck_def}.
 
\section{\label{sec:sum}Summary}

The results of the Lie-transform analysis leading to higher-order guiding-center Hamiltonian dynamics yield the guiding-center Hamiltonian
\begin{eqnarray}
H_{\rm gc} &=& J \left(\Omega + {\cal S}\right) \;+\; |{\bf P}_{0}|^{2}/2m \nonumber \\
 &&+\; e\,\Phi \;-\; \nabla\bdot\left(\mathbb{Q}_{\rm gc}\bdot{\bf E}\right)
 \label{eq:Ham_gc_sum}
\end{eqnarray}
and the guiding-center Lagrangian
\begin{eqnarray}
L_{\rm gc} &=& \left[\frac{e}{c}{\bf A} \;+\; {\bf P}_{0} \;-\; J \left({\bf R} + \frac{1}{2}\,\nabla\btimes\bhat\right)\right]\bdot\dot{\bf X} \nonumber \\
 &&+\; J\;\dot{\theta} \;-\; H_{\rm gc},
 \label{eq:Lag_gc_sum}
 \end{eqnarray}
 where ${\bf P}_{0} \equiv P_{\|}\,\bhat + {\bf E}\btimes e\bhat/\Omega$ and the symmetric dyadic tensor \eqref{eq:Qgc_def} generates a finite-Larmor-radius (FLR) correction to the electrostatic potential energy $e\,\Phi$ in Eq.~\eqref{eq:Ham_gc_sum}. The presence of the gyrogauge fields $({\cal S},{\bf R})$ guarantees that the guiding-center Hamiltonian dynamics derived from the guiding-center Lagrangian \eqref{eq:Lag_gc_sum} are gyrogauge invariant (i.e., these equations are not only gyroangle-independent, but they are also independent on how the gyroangle is measured). 
 
 Next, the explicit dependence of the the guiding-center Lagrangian \eqref{eq:Lag_gc_sum} on the electric and magnetic fields $({\bf E},{\bf B})$ yields guiding-center polarization and magnetization effects, represented by the vector fields $(\vb{\cal P}_{\rm gc},\vb{\cal M}_{\rm gc})$, in the guiding-center Maxwell equations \eqref{eq:gc_QN}-\eqref{eq:gc_Maxwell}. The guiding-center Vlasov-Maxwell equations are, then, derived by an Eulerian variational principle from which exact energy-momentum conservation laws are derived. Here, the guiding-center variational principle \eqref{eq:A_gc} omitted the electric energy density from the guiding-center Lagrangian density, which explicitly guaranteed quasineutrality \eqref{eq:gc_QN} and removed the displacement current density from the guiding-center Maxwell equation \eqref{eq:gc_Maxwell}. In particular, the explicit guiding-center quasineutrality condition removes spurious high-frequency plasma oscillations that violate the guiding-center ordering $\omega \ll \Omega$, while additional spurious high-frequency modes are discussed in App.~\ref{sec:linear_gcVM}.
 
 Future work will explore the the invariance properties of the energy-momentum conservation laws as well as the symmetry properties of the guiding-center stress tensor $\mathbb{T}_{\rm gc}$ defined in Eq.~\eqref{eq:Tgck_def}. In addition, the Hamiltonian formulation of the guiding-center Vlasov-Maxwell equations will be constructed that generalizes the work of Brizard and Tronci \cite{Brizard_Tronci_2016}.

\vspace*{0.2in}

\acknowledgments

The Author wishes to thank one of the referees for raising concerns about spurious high-frequency modes in a set of guiding-center Vlasov-Maxwell equations that does not explicitly impose quasineutrality, and Jeff Candy for a discussion of how gyrokinetic codes handle spurious high-frequency modes in electrostatic gyrokinetic numerical simulations. The present work was supported by the National Science Foundation grant PHY-2206302.

\vspace*{0.1in}

\noindent
{\bf AUTHOR DECLARATIONS}

\vspace*{0.1in}

\noindent
{\bf Conflict of Interest} 

\vspace*{0.1in}

The author has no conflicts to disclose.

\vspace*{0.1in}

\noindent
{\bf Author Contributions}

Alain J. Brizard: Conceptualization (lead); Formal analysis (lead);
Writing – original draft (lead);Writing – review \& editing (lead).

\vspace*{0.1in}

\noindent
{\bf DATA AVAILABILITY}

No data was generated in the course of this work.

\vspace*{0.2in}

\appendix

\section{\label{sec:gyrogauge}Time-dependent Gyrogauge Invariance}

By introducing the fixed unit-vectors $(\wh{\sf 1}, \wh{\sf 2}, \bhat \equiv \wh{\sf 1}\btimes\wh{\sf 2})$, we write the definitions for the rotating unit-vectors
\begin{equation}
\left. \begin{array}{rcl}
\wh{\rho} & \equiv & \cos\theta\;\wh{\sf 1} \;-\; \sin\theta\;\wh{\sf 2} \\
 &  & \\
\wh{\bot} & \equiv & -\;\sin\theta\;\wh{\sf 1} \;-\; \cos\theta\;\wh{\sf 2}
\end{array} \right\},
\label{eq:rho_bot_def}
\end{equation}
where the gyroangle $\theta$ is measured {\it clockwise} from the $\wh{\sf 1}$-axis (for a positively-charged particle), so that $\wh{\bot} \equiv \partial\wh{\rho}/\partial\theta$. We note that, while the choice of the fixed unit-vectors $(\wh{\sf 1},\wh{\sf 2})$ can be made arbitrarily in the plane locally perpendicular to $\bhat$, we must ensure that the resulting guiding-center equations of motion do not depend on a specific choice. Hence, our guiding-center Hamiltonian theory must be {\it gyrogauge}-invariant in the following sense. 

First, we allow the rotation of the unit-vectors $(\wh{\sf 1},\wh{\sf 2})$ about the magnetic unit-vector $\bhat$ by an arbitrary angle $\psi({\bf x},t)$ that depends on the field position ${\bf x}$ at time $t$, so that 
\begin{equation}
\left( \begin{array}{c}
\wh{\sf 1}^{\prime} \\
\wh{\sf 2}^{\prime}
\end{array} \right) \;=\; \left( \begin{array}{cc}
\cos\psi & \sin\psi \\
-\sin\psi & \cos\psi
\end{array} \right) \cdot \left( \begin{array}{c}
\wh{\sf 1} \\
\wh{\sf 2}
\end{array} \right).
\label{eq:gyrogauge}
\end{equation}
Second we require that the rotating unit-vectors \eqref{eq:rho_bot_def} be invariant under this rotation, i.e., $\wh{\rho}^{\prime} = \wh{\rho}$ and $\wh{\bot}^{\prime} = \wh{\bot}$, which implies that the gyroangle $\theta$ must transform as 
\begin{equation}
\theta^{\prime}(\theta,{\bf x},t) \;=\; \theta + \psi({\bf x},t)
\label{eq:theta_prime}
\end{equation}
under the gyrogauge rotation \eqref{eq:gyrogauge}.

Third, we introduce the gyrogauge vector field 
\begin{equation}
{\bf R} \;\equiv\; \nabla\wh{\bot}\bdot\wh{\rho} \;=\; \nabla\wh{\sf 1}\bdot\wh{\sf 2}, 
\label{eq:gyrogauge_R}
\end{equation}
and the gyrogauge scalar field
\begin{equation}
{\cal S} \;\equiv\; \pd{\wh{\bot}}{t}\bdot\wh{\rho} \;=\; \pd{\wh{\sf 1}}{t}\bdot\wh{\sf 2},
\label{eq:gyrogauge_S}
\end{equation}
which transform as ${\bf R}^{\prime} = {\bf R} + \nabla\psi$ and ${\cal S}^{\prime} = {\cal S} + \partial\psi/\partial t$ under the gyrogauge rotation \eqref{eq:gyrogauge}. We, therefore, readily see that a gyrogauge-invariant guiding-center theory can only include the gyrogauge fields $({\bf R},{\cal S})$ either as the one-form $\exd\theta - {\bf R}\bdot\exd{\bf x} - {\cal S}\,\exd t$, the gradient operator $\nabla + {\bf R}\;\partial/\partial\theta$, or the partial time derivative $\partial/\partial t + {\cal S}\,\partial/\partial\theta$, which are all gyrogauge invariant. Finally, we note that the vector fields \cite{RGL_1981}
\begin{eqnarray}
\nabla\btimes{\bf R} &=& -\,\frac{1}{2}\,\epsilon_{ijk}\,b^{i}\;\nabla b^{j}\btimes\nabla b^{k}, \nonumber \\
 && \label{eq:RS_gauge} \\
\pd{\bf R}{t} - \nabla{\cal S} &=& -\;\nabla\bhat\bdot\left(\bhat\btimes\pd{\bhat}{t}\right) \nonumber
\end{eqnarray}
are manifestly gyrogauge independent, since they are expressed entirely in terms of $\bhat$, $\nabla\bhat$, and $\partial\bhat/\partial t$.

Finally, we note that since the magnetic unit vector $\bhat \equiv \partial{\bf x}/\partial s$ is defined as the rate of change of the position ${\bf x}$ of a point as it moves along a magnetic-field line (with $s$ defining distance along that line), we may choose the perpendicular unit vectors $\wh{\sf 1} \equiv \kappa^{-1}\partial\bhat/\partial s$ (where $\kappa$ denotes the Frenet-Serret curvature \cite{Brizard_2015}) and $\wh{\sf 2} \equiv \kappa^{-1}\bhat\btimes\partial\bhat/\partial s$. Hence, the parallel component of the gyrogauge vector field
\begin{equation}
\bhat\bdot{\bf R} \;=\; \pd{\wh{\sf 1}}{s}\bdot\wh{\sf 2} \;=\; \kappa^{-2}\;\frac{\partial^{2}\bhat}{\partial s^{2}}\bdot\left(\bhat\btimes\pd{\bhat}{s}\right) \equiv \tau
\end{equation}
is expressed in terms of the Frenet-Serret torsion $\tau$.

\section{\label{sec:linear_gcVM}Linearized Guiding-center Vlasov-Maxwell Equations}

In this Appendix, we show how the standard linear finite-beta electromagnetic gyrokinetic equations \cite{HLB_1988,Brizard_Hahm_2007} appear as a subset of our guiding-center Vlasov-Maxwell equations. For this purpose, we consider the linearized guiding-center Vlasov-Maxwell equations derived for a uniform background magnetic field ${\bf B}_{0} = B_{0}\,\wh{\sf z}$ and a uniform background Maxwellian distribution $F_{0}$ (with density $N_{0} = \int_{P} F_{0}$ and temperatures $N_{0}T_{\|0} = \int_{P} F_{0}\, mv_{\|}^{2}$ and $N_{0}T_{\bot 0} = \int_{P} F_{0}\,\mu B_{0}$) for each particle species. Consistent with previous gyrokinetic models \cite{Brizard_Hahm_2007,HLB_1988}, a vanishing background electric field is considered, so that the electric and magnetic fields are expressed as
\begin{eqnarray}
{\bf E} &=& -\,\epsilon_{\delta} \left(\nabla\delta\Phi \;+\; \wh{\sf z}\;c^{-1}\partial\delta A_{\|}/\partial t\right), \\
{\bf B} &=& B_{0}\,\wh{\sf z} \;+\; \epsilon_{\delta}\,\nabla\delta A_{\|}\btimes\wh{\sf z},
\end{eqnarray}
where $\|$ and $\bot$ refer to parallel and perpendicular directions with respect to the $z$-axis and $\epsilon_{\delta}$ is a dimensionless ordering parameter associated with the perturbation potentials $(\delta\Phi,\delta A_{\|})$. The guiding-center Vlasov-Maxwell equations \eqref{eq:gcVlasov} and \eqref{eq:gc_QN}-\eqref{eq:gc_Maxwell} will now be linearized up to first order in $\epsilon_{\delta}$ and the gyrogauge contributions are unimportant in what follows since they are not irrelevant in the case of a uniform magnetic field.

First, the guiding-center momentum is expressed as
\begin{equation}
{\bf P}_{0} \;=\; P_{\|}\,\wh{\sf z} \;+\; \epsilon_{\delta}\,\frac{e\,\wh{\sf z}}{\Omega_{0}}\btimes\nabla\delta\Psi,
\end{equation}
where $\delta\Psi \equiv \delta\Phi - (v_{\|}/c)\,\delta A_{\|}$ is the standard electromagnetic effective potential \cite{HLB_1988}, so that the guiding-center velocity \eqref{eq:xdot_P0} becomes
\begin{equation}
\dot{\bf X} \;=\; v_{\|}\,\wh{\sf z} \;+\; \frac{\epsilon_{\delta}\,c}{B_{0}}\left(\wh{\sf z}\btimes\nabla\delta\Psi \;-\; \frac{1}{\Omega_{0}}\frac{d_{0}}{dt} \nabla_{\bot}\delta\Psi\right),
\label{eq:Xgc_dot_App}
\end{equation}
where $d_{0}/dt \equiv \partial/\partial t + v_{\|}\wh{\sf z}\bdot\nabla$. Here, we note that, whenever a term is multiplied by the factor $1/\Omega_{0}$, its contribution comes mainly from ions because of their larger masses compared to electrons. The parallel guiding-center force equation \eqref{eq:Pgc_dot}, on the other hand, is
\begin{equation}
\dot{P}_{\|} \;=\; -\,\epsilon_{\delta}\,e \left(\wh{\sf z}\bdot\nabla\delta\Phi + c^{-1}\partial\delta A_{\|}/\partial t\right),
\label{eq:Pgc_dot_App}
\end{equation}
while the guiding-center Jacobian is
\begin{equation}
{\cal J}_{\rm gc} \;=\; \frac{e}{c}\,B_{\|}^{*} \;=\; \frac{e}{c}\,B_{0} \left( 1 \;+\; \frac{\epsilon_{\delta}c}{B_{0}\Omega_{0}}\;\nabla_{\bot}^{2}\delta\Psi\right),
\label{eq:Jac_App}
\end{equation}
which guarantees that the guiding-center Liouville equation is satisfied up to $\epsilon_{\delta}^{2}$. Using Eqs.~\eqref{eq:Xgc_dot_App}-\eqref{eq:Pgc_dot_App}, the linearized guiding-center Vlasov equation \eqref{eq:gcVlasov} is
\begin{equation}
\frac{d_{0}\delta F}{dt} \;=\; e\left( \wh{\sf z}\bdot\nabla\delta\Phi + c^{-1}\partial\delta A_{\|}/\partial t\right)\;\pd{F_{0}}{P_{\|}},
\label{eq:Vlasov_App}
\end{equation}
which follows from the assumption of a uniform background magnetized plasma. 

Second, taking into account the guiding-center Jacobian \eqref{eq:Jac_App}, the guiding-center charge density is
\begin{equation}
\varrho_{\rm gc} \;=\; \epsilon_{\delta}\;\chi_{\rm gc}\;\nabla_{\bot}^{2}\delta\Phi \;+\; \epsilon_{\delta}\int_{P} e\,\delta F,
\end{equation}
where $\int_{P} v_{\|}\,F_{0} = 0$, and the background Jacobian $eB_{0}/c$ is incorporated into $\int_{P}$. Here, the dominant contribution from the first term (with $4\pi\,\chi_{\rm gc} = c^{2}/v_{\rm A}^{2} = 4\pi N_{0}m_{i}c^{2}/B_{0}^{2}$) comes from ions (because of their larger masses compared to electrons) while the dominant contribution from the second term will come from electrons. In what follows, we will also need the parallel component of the guiding-center current density
\begin{equation}
J_{\|{\rm gc}} \;=\; -\,\epsilon_{\delta}\,\frac{c\beta}{4\pi}\,\nabla_{\bot}^{2}\delta A_{\|} \;+\; \epsilon_{\delta}\,\int_{P} e\,v_{\|}\,\delta F,
\end{equation}
where we assume equal background temperatures $T_{\|0} = T_{\bot0}$, with $\beta \equiv 4\pi N_{0}T_{0}/B_{0}^{2}$, and the dominant contribution from the second term will once again come from electrons. The perpendicular guiding-center current density, on the other hand, is expressed as
\begin{equation}
{\bf J}_{\bot{\rm gc}} \;=\; -\,\epsilon_{\delta}\nabla_{\bot}\left(  \chi_{\rm gc}\;\pd{\delta\Phi}{t} - \frac{c\,\beta}{4\pi}\,\wh{\sf z}\bdot\nabla\delta A_{\|}\right),
\end{equation}
which has no contribution from the perturbed Vlasov distribution $\delta F$ at first order in $\epsilon_{\delta}$ in a uniform magnetized plasma, so that the guiding-center charge conservation law yields
\begin{equation}
\pd{\varrho_{\rm gc}}{t} + \nabla\bdot{\bf J}_{\rm gc} \;=\; \epsilon_{\delta}\,e\int_{P} \frac{d_{0}\delta F}{dt} \;=\; 0,
\end{equation}
where the right side vanishes because of Eq.~\eqref{eq:Vlasov_App}.

Third, up to first order in $\epsilon_{\delta}$, the guiding-center polarization \eqref{eq:Pol_gc} is expressed as
\begin{equation}
\vb{\cal P}_{\rm gc} \;=\; -\,\epsilon_{\delta}\;\frac{\wh{\sf z}}{\Omega_{0}}\btimes\nabla\left( \chi_{\rm gc}\;\pd{\delta\Phi}{t} - \frac{c\,\beta}{4\pi}\,\wh{\sf z}\bdot\nabla\delta A_{\|}\right),
\label{eq:Pgc_App}
\end{equation}
while the guiding-center magnetization \eqref{eq:Mag_gc} is expressed as
\begin{eqnarray}
\vb{\cal M}_{\rm gc} &=& -\;\frac{\beta}{4\pi} \left[ \wh{\sf z} \left( B_{0} + \frac{\epsilon_{\delta}c}{\Omega_{0}}\;\nabla_{\bot}^{2}\delta\Phi\right) \;+\; \epsilon_{\delta}\,\nabla\delta A_{\|}\btimes\wh{\sf z}\right] \nonumber \\
 &&-\; \epsilon_{\delta}\;\frac{c\,\beta}{4\pi\, \Omega_{0}}\;\nabla_{\bot}\left( \wh{\sf z}\bdot\nabla\Phi \;-\; \frac{1}{c}\,\pd{\delta A_{\|}}{t}\right).
 \label{eq:Mgc_App}
 \end{eqnarray}
Here, only ions contribute to the guiding-center polarization \eqref{eq:Pgc_App}, while both ions and electrons contribute to the finite-beta term $-\,\beta\,{\bf B}/4\pi$ in the guiding-center magnetization, with the remaining magnetization is contributed by ions.

From Eqs.~\eqref{eq:Pgc_App}-\eqref{eq:Mgc_App}, we now construct the guiding-center electromagnetic fields ${\bf D}_{\rm gc} = {\bf E} + 4\pi\,\vb{\cal P}_{\rm gc}$ and ${\bf H}_{\rm gc} = {\bf B} - 4\pi\,\vb{\cal M}_{\rm gc}$ to be substituted into the guiding-center Maxwell equations \eqref{eq:gc_QN}  and \eqref{eq:gc_Maxwell}. First, we see that the guiding-center polarization charge density $-\,\nabla\bdot\vb{\cal P}_{\rm gc}$ vanishes in a uniform background magnetized plasma, so that the guiding-center quasineutrality condition \eqref{eq:gc_QN} becomes 
\begin{equation}
\nabla\bdot\vb{\cal P}_{\rm gc} \;=\; 0 \;=\; \varrho_{\rm gc}. 
\label{eq:Poisson_App}
\end{equation}
The parallel component of the guiding-center Maxwell equation \eqref{eq:gc_Maxwell}, on the other hand, yields 
\begin{equation}
\wh{\sf z}\bdot\nabla\btimes{\bf H}_{\rm gc} \;=\; (4\pi/c)\,J_{\|{\rm gc}}, 
\label{eq:Maxwell_par_App}
\end{equation}
where the parallel component of the guiding-center polarization \eqref{eq:Pgc_App} is zero, so that to the parallel guiding-center polarization current vanishes $\wh{\sf z}\bdot\partial\vb{\cal P}_{\rm gc}/\partial t = 0$.

Finally, using the Fourier representation $\delta\chi = \delta\wt{\chi}\,\exp(i{\bf k}\bdot{\bf x} - i\omega t)$, we now derive the linear dispersion relation for the linearized guiding-center Vlasov-Maxwell equations \eqref{eq:Vlasov_App} and \eqref{eq:Poisson_App}-\eqref{eq:Maxwell_par_App} derived for a uniform magnetized plasma. If we assume that only the electron contributions are dominant for the guiding-center charge and parallel current densities (in the limit $\omega \equiv \omega_{{\rm p}e}/\nu \gg k_{\|}
v_{{\rm th}e}$), we find
\begin{equation}
-\,4\pi\,e\int_{P} \left(\begin{array}{c}
\delta\wt{F}_{e} \\
v_{\|}\,\delta\wt{F}_{e}
\end{array}\right)  \;=\; \left(\begin{array}{c}
k_{\|}\,\nu^{2} \\
\omega\,\nu^{2}
\end{array}\right) \left(k_{\|}\,\delta\wt{\Phi} \;-\; \frac{\omega}{c}\,\delta\wt{A}_{\|}\right).
\label{eq:electron}
\end{equation}
Hence, the guiding-center quasineutrality condition \eqref{eq:Poisson_App} becomes
\begin{equation}
\left( k_{\bot}^{2}\,\frac{c^{2}}{v_{\rm A}^{2}} \;-\; k_{\|}^{2}\,\nu^{2}\right)\delta\wt{\Phi} \;=\; -\,\left(\frac{k_{\|}\omega_{{\rm p}e}}{c}\right)\;\nu\;\delta\wt{A}_{\|},
\label{eq:Poisson_DR}
\end{equation}
 while the parallel component of the guiding-center Maxwell equation \eqref{eq:Maxwell_par_App} becomes
\begin{equation}
\left( k_{\bot}^{2} \;+\; \frac{\omega_{{\rm p}e}^{2}}{c^{2}}\right)\delta\wt{A}_{\|} \;=\; \left(\frac{k_{\|}\omega_{{\rm p}e}}{c}\right)\;\nu\;\delta\wt{\Phi},
\label{eq:Maxwell_DR}
\end{equation}
where finite-beta effects coming from the parallel magnetization and guiding-center current densities have cancelled out under the assumption of temperature isotropy. 

Next, we introduce the normalization $K^{2} = k_{\bot}^{2}\rho_{\rm s}^{2}$, $\kappa = k_{\|}/k_{\bot}$, and $(k_{\|}\omega/c)\rho_{\rm s}^{2} = \sqrt{\sigma}\,\kappa\,K/\nu$, where $\sigma = v_{{\rm th}e}^{2}/
v_{\rm A}^{2} = \beta_{e}\,(m_{i}/m_{e})$ denotes a finite-beta parameter and $\rho_{\rm s}^{2} = T_{e}/(m_{i}\Omega_{i}^{2})$, so that the coupled equations \eqref{eq:Poisson_DR}-\eqref{eq:Maxwell_DR} for $(\delta\wt{\Phi},\delta\wt{A}_{\|})$ become
\begin{eqnarray}
K^{2} \left( \frac{c^{2}}{v_{\rm A}^{2}}- \kappa^{2}\,\nu^{2}\right)\delta\wt{\Phi} &=& -\;\sqrt{\sigma}\;\kappa\,K\;\nu\;\delta\wt{A}_{\|}, \nonumber \\
 && \label{eq:dispersion} \\
\left( K^{2} + \sigma\right)\delta\wt{A}_{\|} &=& \sqrt{\sigma}\;\kappa\,K\;\nu\;\delta\wt{\Phi}. \nonumber
\end{eqnarray}
In the electrostatic limit $\sigma \rightarrow 0$ (i.e., $\delta\wt{A}_{\|} \rightarrow 0$), we obtain the electrostatic dispersion relation $\kappa^{2}\nu^{2} = c^{2}/v_{\rm A}^{2}$, which yields the well-known electrostatic H-mode frequency $\omega = \kappa\,(v_{\rm A}\omega_{{\rm p}e}/c) = (k_{\|}/k_{\bot})\,(m_{i}/m_{e})^{\frac{1}{2}}\Omega_{i}$, discovered in electrostatic gyrokinetic models \cite{Lee_1983,Dubin_1983,Lee_1987,Krommes_1993,Zonta_2021}. While this potentially high-frequency mode may violate the fundamental ordering \eqref{eq:ordering} used to derive the guiding-center Vlasov-Maxwell equations, depending on the magnitude of $(k_{\|}/k_{\bot})(m_{i}/m_{e})^{\frac{1}{2}}$, we will now show that this spurious mode disappears when finite-beta electromagnetic gyrokinetic effects are included (see, for example, Lee {\it et al.} \cite{Lee_2001} or Belli and Hammett \cite{Belli_2005}). 

Indeed, when finite-beta ($\sigma > 1$) electromagnetic effects are retained, the coupled equations \eqref{eq:dispersion} yield the dispersion relation 
\[ \frac{c^{2}}{v_{\rm A}^{2}} \;=\; \kappa^{2}\,\nu^{2} \left( 1 \;-\; \frac{\sigma}{K^{2} + \sigma}\right) \;=\; \frac{\kappa^{2}\,\nu^{2}\,K^{2}}{K^{2} + \sigma}, \]
which yields the low-frequency gyrokinetically-modified shear Alfven wave \cite{Lee_2001,Belli_2005,Shi_2015} 
\begin{equation}
\omega \;=\; \frac{k_{\|}\,v_{\rm A}}{1 \;+\; k_{\bot}^{2}\rho_{s}^{2}\,(m_{e}/m_{i}\beta_{e})}, 
\label{eq:shear_Alfven}
\end{equation}
where $k_{\bot}^{2}\rho_{s}^{2}\,(m_{e}/m_{i}\beta_{e}) \sim k_{\bot}^{2}\rho_{s}^{2}\,(m_{e}/m_{i})^{\frac{1}{2}}  < k_{\bot}^{2}\rho_{s}^{2}$ for typical of electromagnetic gyrokinetic models \cite{HLB_1988} with finite beta $\beta_{e} \sim (m_{e}/m_{i})^{\frac{1}{2}}$.

\bibliography{gc_extended}

\end{document}